\documentclass[useAMS,usenatbib]{mn2e}

\usepackage{amssymb}
\usepackage{times}
\usepackage{graphicx}

\newcommand{\be}{\begin{equation}}
\newcommand{\e}{\end{equation}}

\newcommand{\bear}{\begin{eqnarray}}
\newcommand{\ear}{\end{eqnarray}}

\def\apj{ApJ}
\def\apjl{ApJL}
\def\apjs{ApJS}

\def\aj{AJ}

\def\mnras{MNRAS}

\def\k{{\bf k}}

\begin{document}
  \title[Light cone effect on 21-cm power spectrum]{Light cone effect on the reionization 21-cm power spectrum}

\author[Datta et al.]
{Kanan K. Datta$^1$\thanks{E-mail: kdatt@astro.su.se},
Garrelt Mellema$^1$\thanks{E-mail: garrelt@astro.su.se},
Yi Mao$^2$,
Ilian T. Iliev$^3$,
Paul R. Shapiro$^2$ and
\newauthor Kyungjin Ahn$^4$\\
$^1$ Department of Astronomy \& Oskar Klein Centre for Cosmoparticle Physics, Stockholm University, Albanova, SE-10691 Stockholm, Sweden.\\
$^2$ Department of Astronomy and Texas Cosmology Center, University of Texas, Austin, TX 78712, USA\\
$^3$ Astronomy Centre, Department of Physics \& Astronomy, Pevensey II Building, University of Sussex, Falmer, Brighton BN1 9QH, UK \\
$^{4}$ Department of Earth Science Education, Chosun University, Gwangju 501-759, Korea}

\maketitle
\date{\today}

\begin{abstract}
Observations  of redshifted 21-cm radiation from neutral hydrogen during the epoch of reionization (EoR) are considered to constitute the most promising tool to probe that epoch. One of the major goals of the first generation of low frequency radio telescopes is to measure the 
3D 21-cm power spectrum.  However, the 21-cm signal could evolve substantially along the line of sight (LOS) direction of an observed 3D volume, since the received  signal from different planes transverse to the LOS originated from different look-back times and could therefore be statistically different. Using numerical simulations we investigate this so-called light cone effect on the spherically averaged 3D 21-cm power spectrum. 
For this version of the power spectrum, we find that the effect 
mostly `averages out' and observe a smaller change in the power spectrum compared to the amount of evolution in the mean 21-cm signal and its rms variations along the LOS direction. Nevertheless, changes up to $\sim 50\%$ at large scales are possible. In general the power is enhanced/suppressed at large/small scales when the effect is included. The cross-over mode below/above which the power is enhanced/suppressed moves toward larger scales as reionization proceeds. When considering the 3D power spectrum we find it to be anisotropic at the late stages of reionization and on large scales.
 The effect is dominated by the evolution of the ionized fraction of hydrogen during reionization and including peculiar velocities hardly changes these conclusions. We present simple analytical models which explain qualitatively all the features we see in the simulations. 
\end{abstract}


\section{Introduction}
The epoch of reionization (EoR), when the first luminous sources 
reionized the neutral hydrogen in the intergalactic medium (IGM), is
currently the frontier of observational astronomy. Observations of the
Cosmic Microwave Background Radiation (CMBR)
\citep{komatsu11,larson11} and high redshift quasar absorption spectra
\citep{becker01,fan06,willott09} jointly suggest that reionization
took place over an extended period spanning the redshift range $6 \le
z \le 15$ \citep[see e.g.][]{2011MNRAS.413.1569M}.
Observations of high redshift Ly $ \alpha $-emitting galaxies 
\citep{malhotra04,ouchi10,kashikawa11} and gamma ray bursts
\citep{totani06} are also consistent with this picture.

Observations of redshifted 21-cm radiation are considered to constitute the most promising tool to probe the EoR \citep[for a review see][]{fur06}. For the past few years substantial efforts have been undertaken both on the theoretical and experimental side \citep[reviewed in][]{morales10}. The first generation of low frequency radio telescopes ( GMRT\footnote{http://gmrt.ncra.tifr.res.in/}, LOFAR\footnote{http://www.lofar.org/}, MWA\footnote{\citet{2009IEEEP..97.1497L}, http://www.mwatelescope.org}, PAPER\footnote{\citet{parson10}}) is either operational or will be operational very soon. Preliminary results from these facilities include foreground measurements at EoR frequencies \citep{ali08,bernardi09,pen09,paciga10} as well as some constraints on reionization \citep{bowman10,paciga10}. 

Motivated by the detection possibility of the EoR 21-cm signal and the subsequent science results, a wide range of efforts are ongoing on the theoretical side with the goal to understand the physics of reionization and its expected 21-cm signal. \citet{fur04} developed analytical models to calculate the ionized bubble size distribution and use this for calculating the 21-cm power spectrum. Such models are  very useful in predicting the signal quickly for a wide range of scales and investigating the large parameter space. However, they cannot incorporate details of reionization  and become less accurate  when the bubbles start overlapping. Numerical simulations are probably the best way to predict the expected 21-cm signal. Although challenging, there has been considerable progresses in simulating the large scale 21-cm signal during the entire EoR \citep{iliev06,mellema06,mcquinn07,shin08,baek09}. More approximate but much faster semi-numerical simulations of the structure and evolution of reionization and the 21-cm signal
have also been developed \citep{zahn07,mesinger07,santos08,geil08,thomas09,choudhury09}. These methods are capable of generating volumes with sizes as large as $\sim 1 \rm {Gpc}^3$ \citep{alvarez09,santos10}. Many aspects such as source properties, feed back effects, distribution and properties of sinks have also been investigated in detail \citep[see][ for a review on reionization simulations]{trac09}. 

\begin{figure*}
\includegraphics[width=.26\textwidth, angle=270]{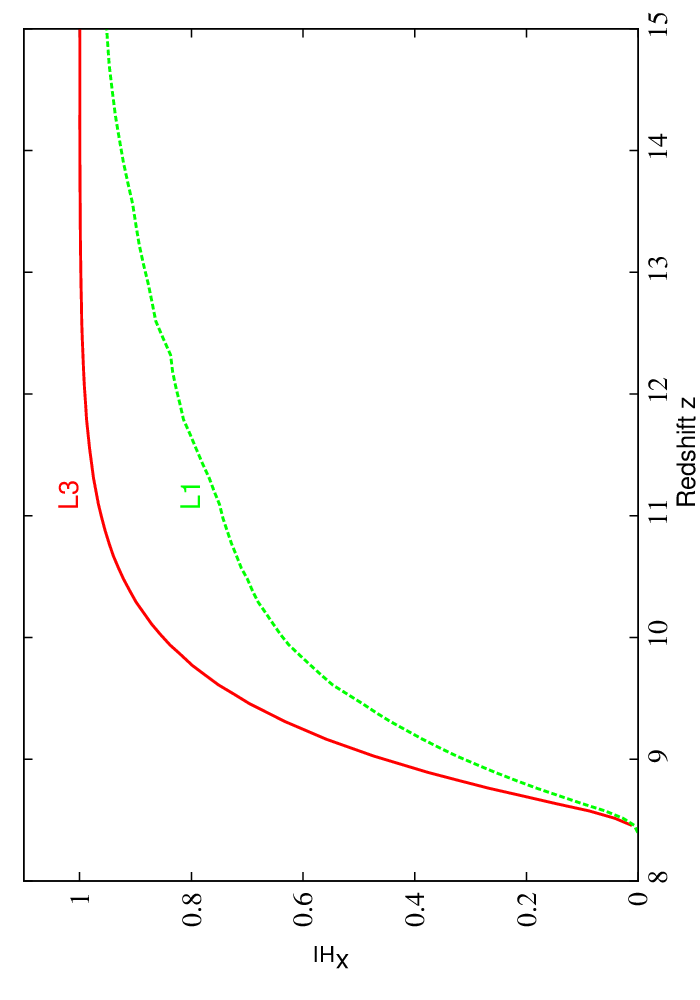}
\includegraphics[width=.26\textwidth, angle=270]{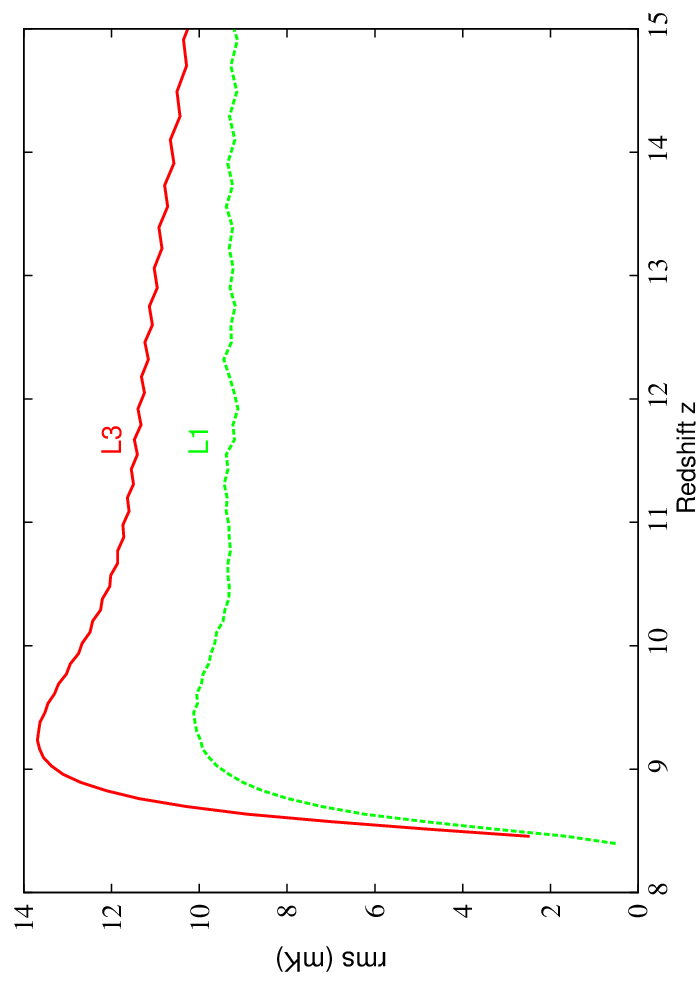}
\caption{The evolution of the mean mass averaged neutral fraction $x_{\rm{H I}}$ and the rms of 21-cm fluctuations with redshift for the two different reionization simulations L1 and L3.}
\label{fig:x_HI_z}
\end{figure*}

One of the major goals of all first generation EoR telescopes is to measure the spherically averaged 3D 21-cm power spectrum.  Measurements of the 21-cm power spectrum will provide a wealth of information about the timing and duration of reionization, large scale distribution of H I and its evolution, source properties and clustering \citep{ali05,sethi05,datta07,lidz08,barkana09}. To obtain the spherically averaged 3D power spectrum one needs to average over the 3D volume produced by the observations. Of this 3D volume one axis (the LOS axis) is along the frequency direction. Since light from the lower frequency side of the 3D volume takes a longer time to reach us than light from the high frequency side,
the observer will see reionization in an earlier phase at the  lower frequency side than at the higher frequency side. The statistics of 21-cm fluctuations could therefore be changing over the observed volume. As we will see in section 2 and 3, in some reionization scenarios the change could be substantial especially near the end of reionization. Almost all previous studies calculate the 3D 21-cm power spectrum without taking this effect into account. In this paper we investigate the effect of LOS evolution or the so called `light cone' effect on the measured 21-cm power spectrum, using numerical simulations to quantify it. Understanding the light cone effect is important because it will be present in the data and needs to be taken into account when interpreting the observed 21-cm power spectrum. Our aim is to understand under which conditions and at what scales this effect needs to be considered. 

The light cone effect is well known from studies of galaxy clustering \citep[see e.g.][]{matsubara97}. In the context of 21-cm studies of reionization it was first considered by \citet{barkana06}. These authors studied analytically the anisotropic structure of the two point correlation function caused by the effect. This appears to be only work that considered the effect of a changing source population. However, more work has been done on the light cone effect for a single bright source, such as a QSO. For this case the effect will make the H II region appear to be teardrop-shaped \citep{wyithe05, yu05, majumdar10}. The effects on the power spectrum and correlation function for this case were investigated by \citet{sethi08}. In addition, for very luminous sources the effect of relativistically expanding H~II regions \citep{shapiro06} would have to be added to the one purely due to evolution of the signal along the LOS.

Bright QSOs are quite rare, so the more common form of the light cone effect will be due to the evolving source population and the growing H~II regions around groups of sources. Our aim is to study this version of the effect on the spherically averaged 3D and the 1D LOS power spectra using realistic numerical simulations of reionization.

The paper is organized  as follows. Section 2 briefly describes our simulations and the procedure used to generate light cone cubes. We present our results in Section 3. Section 4 describes two simple toy models which explain qualitatively the main features we see in the simulation results. Section 5 investigates how the inclusion of peculiar velocities affect our results. We summarize our results and conclusions in Section 6. The cosmological parameters we use throughout the paper are $\Omega_m=0.27, \Omega_k=0, \Omega_b=0.044, h=0.7, n=0.96$ and $\sigma_8=0.8$, consistent with the {\it WMAP} seven-year results \citep{komatsu11}.

\section{Simulation}

\subsection{The redshifted 21-cm signal}

The 21-cm radiation is emitted when neutral hydrogen atoms go through spin-flip transitions. The radiation can be decoupled from the cosmic microwave background (CMB) photons either through collisions with hydrogen atoms and free electrons \citep{purcell56, field59, zygelman05}  or through Ly$\alpha$ photon pumping \citep{wouthuysen52, field59, chen04, hirata06, chuzhoy06}. This makes 21-cm radiation detectable either in emission or absorption against the CMB. The differential brightness temperature with respect to the CMB is commonly written using the spin temperature $T_s$ as
\be
\delta T_b \approx 27.4 \, {x}_{\rm H I} \, \rm{mK}\, \left ( \frac{1+z}{10} \right )^{1/2} \frac{(T_s-T_{CMB})}{T_s}(1+\delta_\mathrm{H})\,
\e
where ${x}_{\rm H I}$ and $\delta_\mathrm{H}$ are the mass averaged neutral fraction and the density fluctuations of hydrogen. Note that the 21-cm signal remains undetectable when the spin temperature $T_s$ is coupled to the CMB temperature $T_{\rm{CMB}}$. During the EoR, $T_s$ is expected to be coupled to the gas kinetic temperature through  Ly$\alpha$ photon coupling. In addition the gas kinetic temperature is expected to be much higher than the CMB temperature due to heating by shocks, X-rays and Ly$\alpha$ photons. This would make the redshifted 21-cm signal visible in emission. We assume here that $T_s \approx T_{\rm{gas}}\gg T_{\rm{CMB}}$ which makes the 21-cm signal independent of the actual value of $T_s$. This is a reasonable assumption during the later stages of the EoR. During the initial stages of reionization, when there are few sources of radiation, this assumption might not hold \citep{baek10,thomas10}. The next subsection describes how we simulate the fluctuations in the H I density.

\begin{figure*}
\includegraphics[width=.4\textwidth, angle=270]{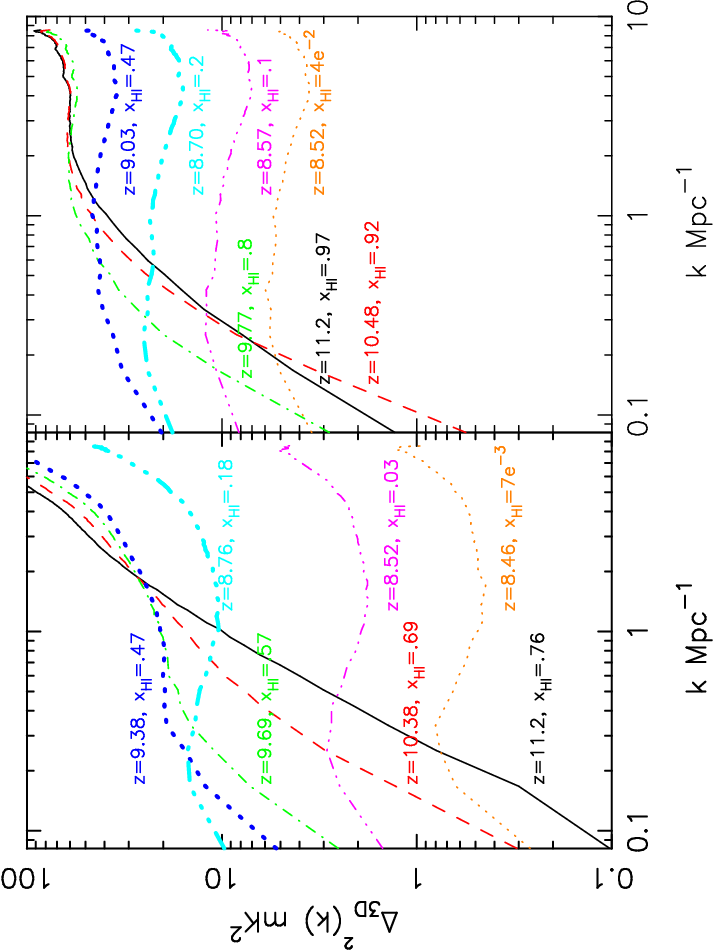}
\caption{The evolution of  dimensionless spherically averaged 3D 21-cm power spectrum $\Delta^2_{\rm{3D}}(k)$ with redshift in the L1 (left panel) and L3 (right panel) simulations. The light cone effect is not taken into account.}
\label{fig:ps3dg}
\end{figure*}
\subsection{N-Body and radiative transfer runs}
Details about our simulation methodology (N-body simulation and the subsequent radiative transfer) have been presented in previous papers \citep{iliev06,mellema06,iliev07,iliev11}. \cite{iliev11} described the simulations we use here in more detail. Here we only present a brief overview of the major features of these simulations. 

We start by simulating the evolution of the dark matter distribution using the CubeP$^3$M N-body code in a comoving volume of $(163 \, \rm{cMpc})^3$ using $3072^3$ particles and $6144^3$ cells. This implies particle masses of $5.5 \times 10^6 M_{\odot}$ and a minimum resolvable halo mass of $\sim 10^8 M_{\odot}$ which approximately matches the minimum mass of haloes able to cool by atomic cooling. The N-body simulations give the DM density field, locations and masses of haloes. We then assume that the baryons trace the DM density field and assign an ionizing photon luminosity to each halo assuming it to be proportional to the halo mass,
\be
{\dot N_{\gamma}}=g_{\gamma}\frac{M_h\Omega_b}{10 \Omega_m m_p}\,,
\e
where ${\dot N_{\gamma}}$ is the number of ionizing photons emitted per time. $M_h$ and $m_p$ are  the halo mass and proton mass respectively. The efficiency parameter $g_{\gamma}$ can be written as,
\be
g_{\gamma}=f_{\gamma}\frac{10 \, \mathrm{Myr}}{\Delta t}\,,
\e
where $f_{\gamma}$ is the number of ionizing photons emitted into the IGM per baryon per star forming episode (which is taken to be the same as simulation time-step, about $11.5\times 10^6$~years). This makes the factor $f_{\gamma}$ the product of the escape fraction of ionizing radiation $f_\mathrm{esc}$, the star formation efficiency $f_\mathrm{*}$, and the number of ionizing photons produced per baryon for a given initial mass function (IMF), $N_{\gamma}$. The latter number is around 5,000 for a Salpeter IMF and $\sim 10$ times higher for top heavy IMFs, and the two fractions are of the order 10\%. This gives $f_{\gamma}$ values in the range 1 to 100. We divide the halos into low mass atomically cooling halos of mass $10^8$--$10^9$~M$_{\odot}$ and high mass atomically cooling halos of masses higher than that. To take into account feedback on the low mass halos, we turn off their ionizing luminosity when they are located in an ionized region. 

We then calculate the transfer of ionizing photons with the C$^2$-Ray code \citep{2006NewA...11..374M} on a $256^3$ grid. Ionizing photons are traced from every source cell to every grid cell within a given time step. This gives us the distribution of the H I fraction in the volume at different redshifts.

Here we consider three cases for the ionizing photon luminosity. In the first simulation, labeled L1 in \citet{iliev11}, $g_{\gamma}=8.7$ or $f_{\gamma}=10$ for sources of mass $>10^9$ M$_\odot$ and $g_{\gamma}=130$ or $f_{\gamma}=150$ for sources of mass between $10^8 - 10^9$ M$_\odot$. In the second simulation, labeled L2, these factors are $g_{\gamma}=1.8$ or $f_{\gamma}=2$ and $g_{\gamma}=8.7$ or $f_{\gamma}=10$. In the third case, labelled L3, sources of mass below $2.2 \times 10^9$ M$_\odot$ have been turned off. To end reionization at almost the same time as L1, the active sources have been assigned a higher luminosity with $g_{\gamma}=21.7$ or $f_{\gamma}=25$. This particular setup was chosen to make it analogous to the older simulations from \citet{iliev08}, but updated for the WMAP5 cosmology.

The simulations L1 and L3 are the ones which have the strongest evolution, where L3 due to the lack of low mass sources has the fastest evolution. Since the light cone effect is caused by evolution, below we will focus on these two cases and briefly mention the more slowly evolving case L2 at the end of Section~\ref{sect:sim_results}.

\begin{figure*}
\includegraphics[width=.47\textwidth, angle=0]{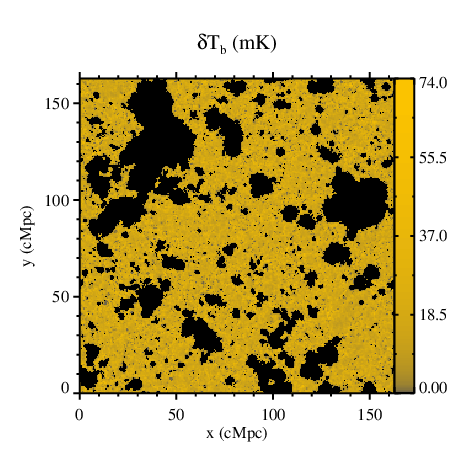}
\includegraphics[width=.47\textwidth, angle=0]{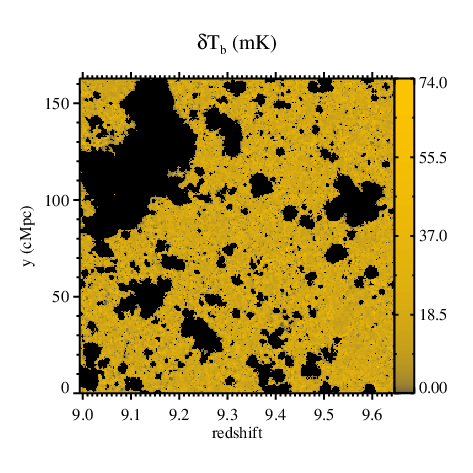}
\caption{Two-dimensional slices of the 21-cm signal from a coeval cube at $z=9.31$ (left) and light cone cube (right) with central redshift $z_\mathrm{c}=9.31$ 
(simulation L1). For the latter the $x$-axis corresponds to the LOS direction. 
Comparison of these two slices shows the effect of evolution on the sizes of H II regions at both the front and back sides of the light cone cube. Note that for this visualization the mean signal has not been subtracted from the light cone cube.}
\label{fig:image}
\end{figure*}

\subsection{Light cone cube}
The simulations provide us with so-called `coeval cubes', 3D volumes of density and H I fraction at the same cosmological redshift. The extent of these cubes corresponds to a redshift range of $\Delta z \approx 0.6 - 0.9$, depending on redshift. An observer can not observe these coeval cubes, but we use them to create observable `light cone' cubes. The procedure which was previously introduced in \citet{mellema06}, is as follows:

\begin{itemize}

\item From the simulation we obtain a set of $N$ coeval 21-cm cubes at
  redshifts $z_1,z_2.....z_N (z_1<z_2.....<z_N)$ each of integer size
  $M^3$ and physical comoving size $L^3$. For the case at hand $M=256$
  and $L=163$~cMpc.
\item Starting at $z_1$ we create a redshift series $z_\mathrm{LC}$ of
  length $m=K\times M$ ($K\le N$) which will constitute the redshift
  (LOS) axis of the light cone `cube' (which will therefore not be
  cubical). Each consecutive redshift in the series is the same
  comoving distance apart, namely $L/M$.
\item We then construct the light cone cube by stepping through this
  redshift series and constructing the 21-cm slices of size $M^2$ for
  each redshift.
\item To create the $p$th 21-cm slice of the light cone cube, we first
  calculate the integer division $p/M$ and its remainder $q$. We
  pick up the $q$th slice from the two coeval cubes at $z_l$ and
  $z_{l+1}$, where $z_l \le z_\mathrm{LC}(p) \le z_{l+1}$, and use
  linear interpolation in redshift to create an 21-cm slice at
  $z_\mathrm{LC}(p)$.

\end{itemize}


We should point out that the light cone cubes constructed this way differ from the observational ones in that the field of view has a constant comoving size and not a constant angular size. This is a natural consequence of the way they are constructed from the simulation results and makes it easier to construct the 3D power spectra from them. For the real interferometric observations the angular field of view would be slowly changing as a function of frequency and the physical comoving size depends on the redshift via the angular-size distance relationship. For determining the 3D 21-cm power spectrum in $k$-space it will always be possible to extract a volume with a constant comoving field of view from the observational data.

From the L1 simulation we extracted 35 coeval cubes at redshifts spanning from $z=11.20$ to $8.4$.  In this model reionization starts earlier with the mass weighted ionization fraction reaching $1 \%$ and $50\%$ around redshifts $z=17.22$ and $9.46$ respectively (see Figure \ref{fig:x_HI_z}). In the L3 model, where the smaller mass halos do not contribute, the reionization process starts later (because massive sources form later) and the $1 \%$ and $50\%$ points are reached around redshifts $z=12$ and $9$ respectively. By construction, the two simulations complete reionization at the same redshift of 8.4, so in the L3 simulation reionization proceeds faster. Because L3 has fewer sources, the characteristic bubble size for a given neutral fraction $\rm {x_{H I}}$ is bigger. Both models are consistent with the recent CMB measurements of the electron scattering optical depth. Figure \ref{fig:x_HI_z} shows the evolution of the mass averaged neutral fraction $\rm {x_{H I}}$ (left panel) and the rms of 21-cm fluctuations (right panel) for the two models. Note that for L3 the rms is higher than for L1 because of the larger ionized bubbles which amplify the rms signal. 

Since the comoving distance between $z=11.20$ to $8.4$ is larger than 163~cMpc our full light cone cube is constructed by using the periodicity of our cosmological volume. However, this does mean that we pass through the same structures several times and our power spectra would be unphysical below scales of $\sim 0.08$~Mpc$^{-1}$. We therefore limit our power spectrum analysis to subvolumes of LOS size 163~cMpc, which roughly corresponds to a frequency depth of $\sim 10$~MHz.

\begin{figure*}
\begin{center}
\includegraphics[width=.6\textwidth, angle=270]{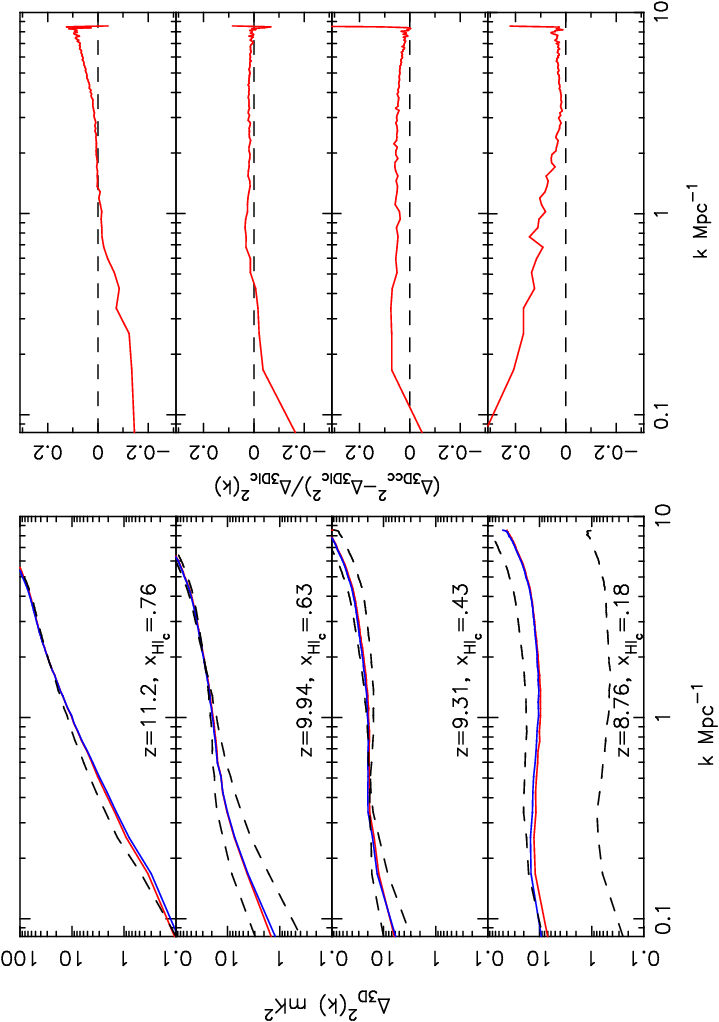}
\caption{The light cone effect on the 21-cm power spectrum $\Delta^2_{\rm{3D}}(k)$ in simulation L1. Left panels: $\Delta^2_{\rm{3D}}(k)$ for the light cone cube (red), coeval cube (blue) centered around the redshift mentioned in the plot and coeval cubes (two dashed lines) for redshifts corresponding to back and front sides. Right panels: the relative difference $(\Delta^2_{\rm{3Dcc}}-\Delta^2_{\rm{3Dlc}})/\Delta^2_{\rm{3Dlc}}$ where `cc' and `lc' stand for coeval and light cone cube respectively. }
\label{fig:ps3de_f10}
\end{center}
\end{figure*}

\begin{figure*}
\begin{center}
\includegraphics[width=.6\textwidth, angle=270]{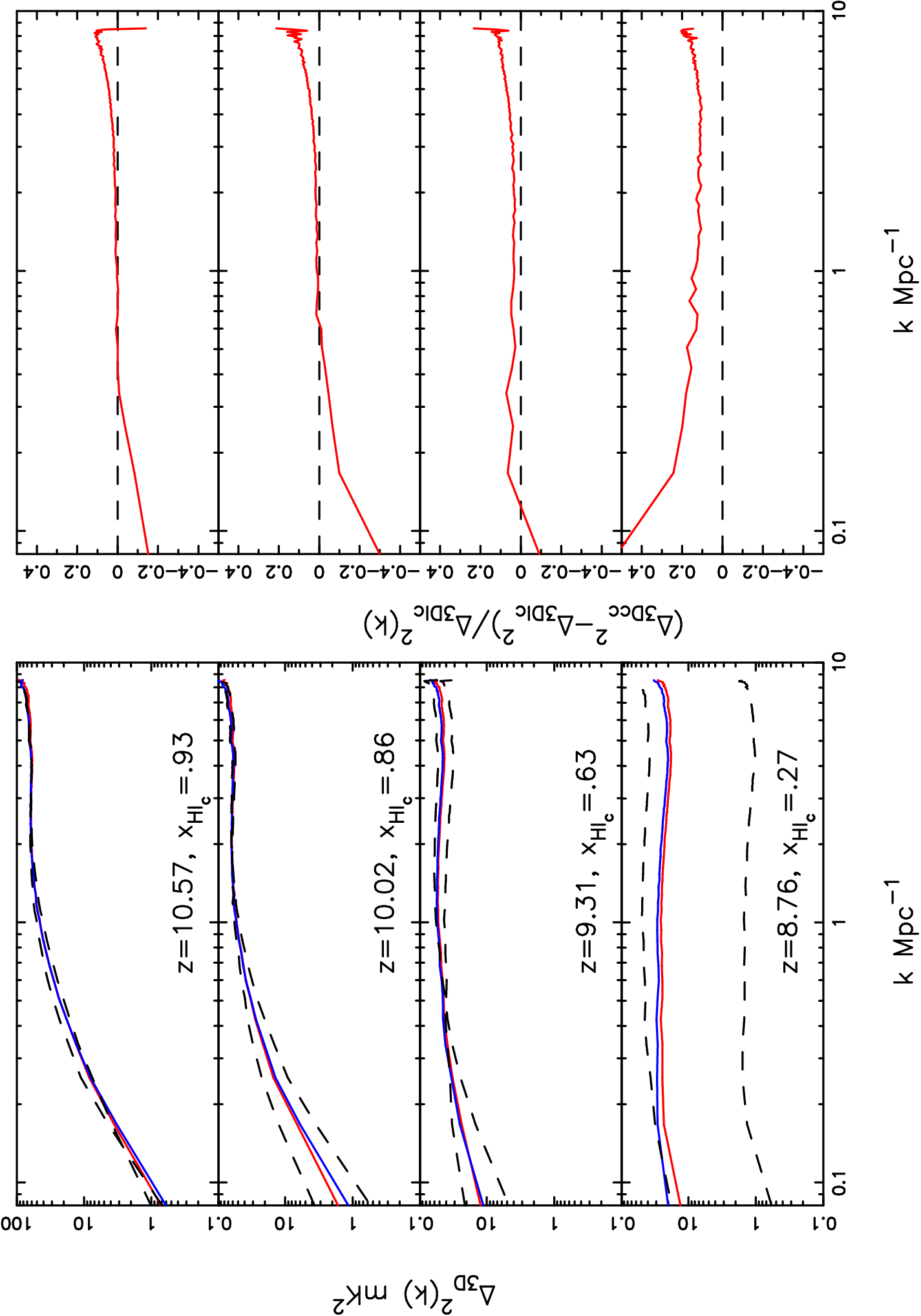}
\caption{Same as Figure \ref{fig:ps3de_f10}, but for simulation L3.}
\label{fig:ps3de_f25}
\end{center}
\end{figure*}

\begin{figure*}
\begin{center}
\includegraphics[width=.35\textwidth, angle=270]{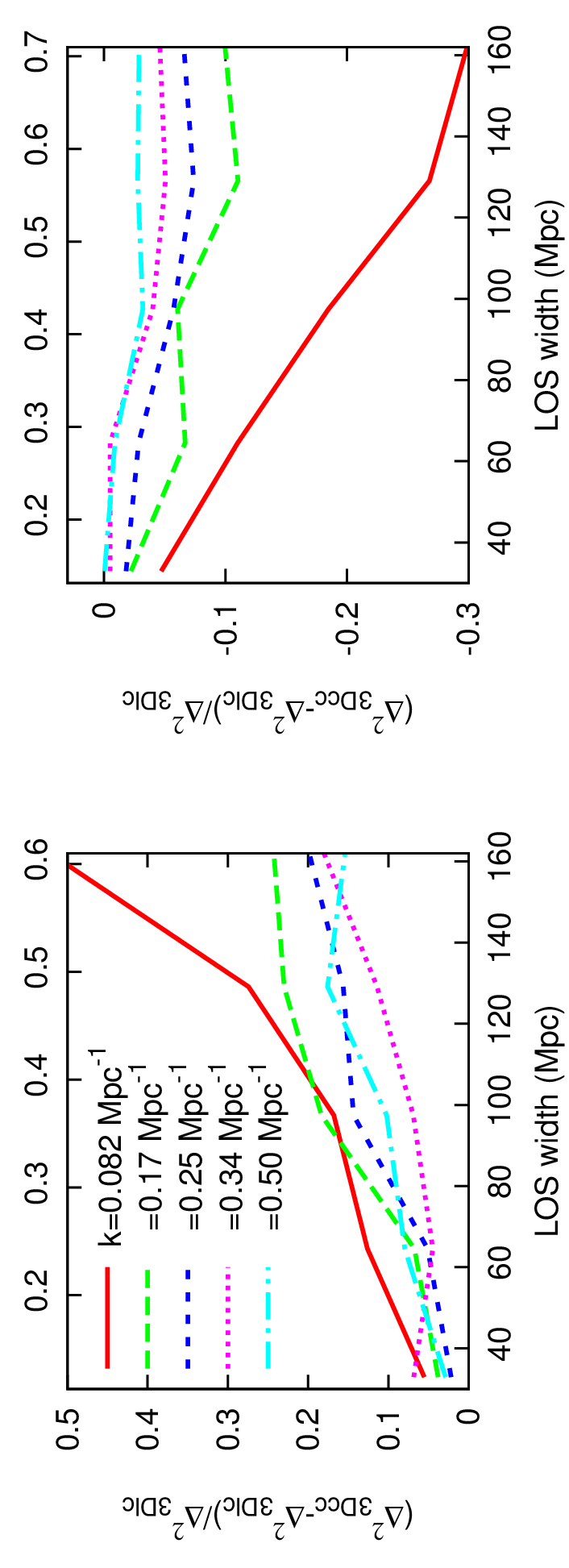}
\caption{Change of the light effect for different LOS widths. The plots show $(\Delta_{\rm 3Dcc}^2-\Delta_{\rm 3Dlc}^2)/\Delta_{\rm 3Dlc}^2$ as a function of LOS width (and $\Delta z $ above the top x-axis) for different  ${\rm k}$ modes at two central redshifts $z_{\rm c}=8.76$ (left panel) and $10.02$ (right panel) for the L3 simulation.}
\label{fig:change-bw}
\end{center}
\end{figure*}

Figure \ref{fig:ps3dg} shows the dimensionless spherically averaged 3D 21-cm power spectra $\Delta_{\rm {3D}}^2(k)=k^3P_{\rm {3D}}(k)/2\pi^2$ \citep{peacock} for coeval cubes at different redshifts for both simulations.  In the beginning of reionization the power spectrum is dominated by the density fluctuations and quite similar to the underlying dark matter power spectrum. As reionization proceeds the growing ionized bubbles add power at larger scales. When reionization reaches $\sim 50 \%$ the power reaches a maximum at larger scales. As the neutral fraction goes down further the overall amplitude of the power spectrum also goes down. Note once again that the L3 simulation has more power than the L1 model because of the larger ionized bubbles. From Figure~\ref{fig:ps3dg} we see the power spectrum evolve both in amplitude and in slope \citep[see][ for a detailed discussion on the power spectrum evolution]{lidz08}. The details of the evolution depend on the reionization scenario (for example, the evolution is much faster in the L3 model). If we consider L1, we can see that the power spectrum  $\Delta_{\rm {3D}}^2(k)$ at $k=0.1 \, \rm {Mpc}^{-1}$ changes from $\sim 0.25 \, \rm {mK}^2 $ to $\sim 10 \, \rm {mK}^2$ in the redshift range $z=8.46$ to $8.89$. Such rapid evolution of the power spectrum (by a factor of $\sim 40$ at $k=0.1 \, \rm {Mpc}^{-1}$ ) within $\Delta z=0.43$ provided the motivation for studying the light cone effect.

\begin{figure*}
\begin{center}
\includegraphics[width=.35\textwidth, angle=270]{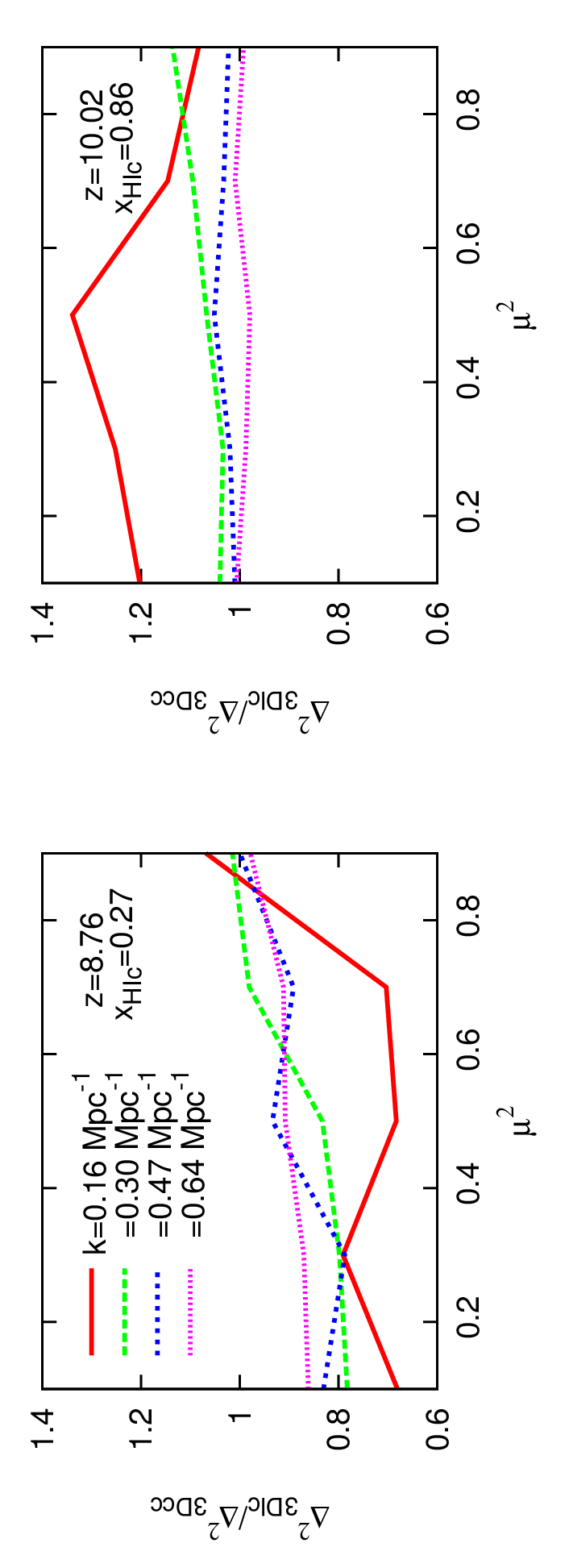}
\caption{Anisotropy of the 3D power spectrum. The plots show the ratio $\Delta^2_{\rm 3Dlc}/\Delta^2_{\rm 3Dcc}$ as a function of $\mu^2$ for different $k$-modes for two central redshifts $z_{\rm c}=8.76$ (left panel) and $10.02$ (right panel) for L3 simulation.}
\label{fig:ps-aniso}
\end{center}
\end{figure*}

\begin{figure*}
\begin{center}
\includegraphics[width=.6\textwidth, angle=270]{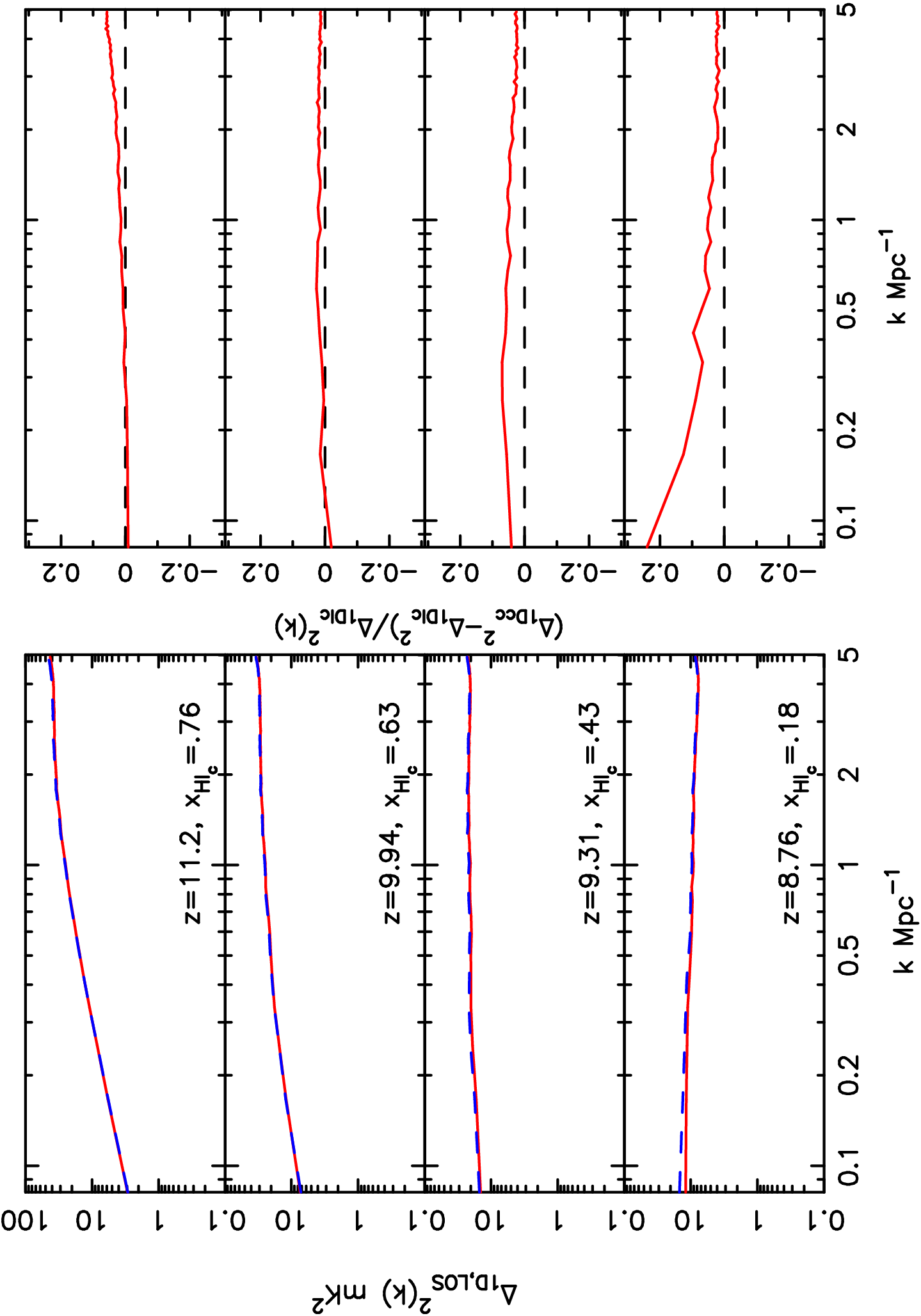}
\caption{Same as Figure \ref{fig:ps3de_f10}, but for $\Delta^2_{\rm{1D}}(k)$ 
and simulation L1.}
\label{fig:ps1de_f10}
\end{center}
\end{figure*}


\section{Effect of evolution}
\label{sect:sim_results}
\subsection{Spherically averaged power spectrum}
\begin{table}
\caption{Details about the simulated cubes from simulation L1, used for our analysis.}
\label{tab:table_L1}
\begin{tabular}{|c|c|c|c|c|}
\hline
$z_c$ & $x_{\rm{H Ic}}$ & Redshift& $x_{\rm{H I}}$& rms\\ 
&&extent&variation&variation\\
&&&&(mK)\\
\hline
8.76 & $0.18$ & $8.46-9.07$ &$7\times10^{-3}-0.33$ & $1.0-9.7$ \\
\hline
9.31 & $0.44$ & $8.99-9.65$ & $0.3-0.55 $ &$9.5-10$ \\
\hline
9.94 & $0.63$ & $9.59-10.3$ & $0.54-0.68$ & $10-9.4$ \\
\hline
11.2 & $0.76$ & $10.8-11.63$ & $0.73-0.8$ & $9.3-9.2$ \\
\hline
\end{tabular}
\end{table}

\begin{table}
\caption{Same as Table \ref{tab:table_L1} for simulation L3.}
\label{tab:table_f25}
\begin{tabular}{|c|c|c|c|c|}
\hline
$z_{\rm{c}}$ & $x_{\rm{H Ic}}$ & Redshift& $x_{\rm{H I}}$& rms\\ 
&&extent&varies from&varies from\\
&&&&(mK)\\
\hline
8.76 & $0.27$ & $8.46-9.07$ &$1\times10^{-2}-0.5$ & $2.5-13.5$ \\
\hline
9.31 & $0.63$ & $8.99-9.65$ & $0.45-0.76 $ &$13.2-13.3$ \\
\hline
10.02&$0.85$ & $9.59-10.3$ & $0.75-0.9$ & $13.3-12.2$ \\
\hline
10.57 & $0.93$ & $10.2-10.96$ & $0.88-0.96$ & $12.4-11.7$ \\
\hline
\end{tabular}
\end{table}

\begin{figure*}
\includegraphics[width=.66\textwidth, angle=0]{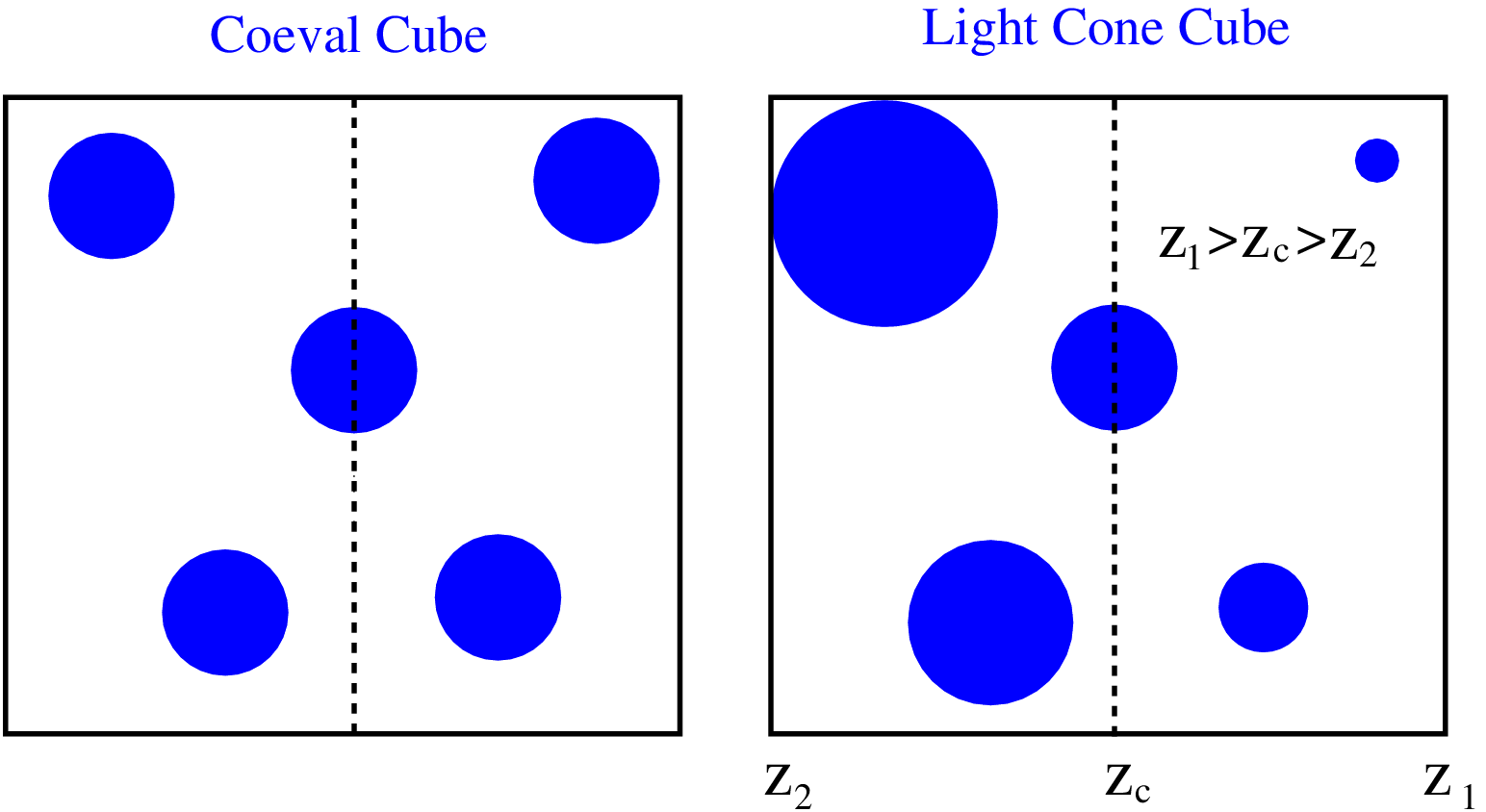}
\caption{Cartoon illustration of Toy model 1. The left panel represents the coeval cube containing ionized bubbles of the same size. The right panel represents the light cone cube with bubbles from back/front side appearing smaller/larger in comparison to the coeval cube.}
\label{fig:toymodel1}
\end{figure*}

In this section we present and discuss our results on the light cone
effect. Figure \ref{fig:image} shows two 21-cm images constructed from
simulation L1, the left one from a coeval cube at $z=9.31$, and the
right one from a light cone cube, where the horizontal axis
corresponds to the LOS direction and the central redshift is $z=9.31$.
We see that ionized regions (black patches) at the higher $z$ side (righthand side) are smaller in the light cone image than in the coeval image. Conversely, the ionized regions at the lower $z$ end (lefthand side) are larger in the light cone cube. 

Figures \ref{fig:ps3de_f10} (for L1) and \ref{fig:ps3de_f25} (for L3) show the effect of evolution on the spherically averaged 3D 21-cm power spectrum.  Note that we do not include the modes ${\bf k} (k_x=0,k_y=0,k_z)$ when we take spherical average over all modes between $k$ and $k+dk$.  Since we are performing this analysis in the context of radio-interferometric measurements that do not measure the modes ${\bf k} (k_x=0,k_y=0,k_z)$, it is appropriate not to include those modes. It is in fact quite important for the analysis we present. Excluding the modes ${\bf k} (k_x=0,k_y=0,k_z)$ changes the 3D power spectrum from the light cone cube considerably. We discuss this more extensively in Appendix A.

Details about the central redshift ($z_c$), mass averaged neutral fraction at the central redshift ($x_{\rm{H Ic}}$), redshift range, neutral fraction range and the range of the rms variations in the 21-cm signal for the different light cone cubes from simulations L1 and L3 are given in Tables \ref{tab:table_L1} and  \ref{tab:table_f25} respectively. The left panels of Figure.~\ref{fig:ps3de_f10} show the spherically averaged 3D power spectra $\Delta^2_{\rm{3D}}(k)$ for light cone cubes at four different central redshifts. For comparison it also shows the power spectra for coeval cubes at three redshifts (high, low redshift end and central redshift of the light cone cube). The righthand panels plot the relative difference in the power spectra between the coeval cube at the central redshift and the light cone cube. General features we see in all panels are that  the effect is stronger on large scales and increases as we go up in scale. In addition, we find that the power is enhanced (suppressed) with respect to the coeval cube at large (small) scales for neutral fractions $x_{\rm{H I}}\lesssim 0.5$. During the last phase of reionization (bottom panels) the power in the light cone cube is suppressed  at all scales. At redshift $z=11.20$ we see that the power spectrum is enhanced by $\sim 15 \%$  at modes $k<0.5 \, \rm{Mpc}^{-1}$ and suppressed by $\sim 10 \%$ at small scales. At redshift $z=9.94$ we do not see much effect except on the largest scale where the power is larger by $\sim 15 \%$. This is rather surprising since the evolution of the 21-cm signal is stronger in this redshift interval than in the $z=11.20$ band.  For the cube centered around redshift $z=9.94$ the neutral fraction and the rms change from $0.547$ to $0.68$ and $10$ to $9.4$~mK and the power spectrum is amplified by a factor of $\sim 7$ at $k=0.1 \rm{Mpc}^{-1}$ (see Table \ref{tab:table_L1} and the last two left panels of Figure \ref{fig:ps3de_f10}), much more than what we see in the cube around  $z=11.20$. So we would expect a larger effect than what we see in Figure \ref{fig:ps3de_f10} at $z=9.94$. This trend continues and we see almost no effect for redshift $z=9.31$ where the neutral fraction and the rms change even more ($0.325 \rightarrow 0.547$ and $9.4 \rightarrow 10$~mK) and the power spectrum is amplified by a factor of $\sim 5$ at $k=0.1$~{Mpc}$^{-1}$. At redshift $z=8.76$, instead of an enhancement we see suppression on all scales with differences up to $30\%$. In the L3 simulation this  suppression is up to $50 \%$ at redshift $z=8.76$. All other features are quite similar in the L3 model (Figure \ref{fig:ps3de_f25}) even though the reionization process proceeds faster and the ionized regions are larger in this model. 

Another way to describe the trend we see is that we find a cross-over mode $k_{\rm{cross-over}}$ below which power is enhanced and above which it is suppressed. The cross-over scales $k_{\rm{cross-over}}$ shifts towards lower $k$ as the reionization proceeds. At the end of reionization the cross-over mode is lower than the lowest mode we measure from the simulation box. 

\subsection{Light cone effect as a function of LOS width}
 
Above we present results using the entire cubes i.e, for a LOS width corresponding to the size of our simulation volume. However, it is interesting to explore how the effect changes as one reduces the LOS width. Obviously in
the limit of small widths, the light cone effect will disappear, so considering
a range a widths allows us to study how it varies with width.

Here we consider sub-boxes of different LOS widths $\Delta z $ and
calculate the quantity $(\Delta_{\rm 3Dcc}^2-\Delta_{\rm
  3Dlc}^2)/\Delta_{\rm 3Dlc}^2$ for different $k$ modes. Figure
\ref{fig:change-bw} shows $(\Delta_{\rm 3Dcc}^2-\Delta_{\rm
  3Dlc}^2)/\Delta_{\rm 3Dlc}^2$ as a function of LOS width (and
$\Delta z $) for different $ k$ modes at two central redshifts $z_{\rm
  c}=8.76$ and $10.02$ for the L3 simulation. As expected, we see that
the quantity $(\Delta_{\rm 3Dcc}^2-\Delta_{\rm 3Dlc}^2)/\Delta_{\rm
  3Dlc}^2$ decreases for smaller LOS width (and $\Delta z $). As
discussed later in the subsection 4.3, we expect the quantity to
increase quadratically with the LOS width. However, due to the smaller 
number of modes available for the smaller $\Delta z$, the results are too
noisy to test this expectation, although they are roughly consistent with
it.

We do not show results for the other two central redshifts of L3 and
L1 simulation as the light cone effect is relatively smaller for these,
but find similar results there. These results suggest that
measurements of the light cone effect for different LOS widths can, in
principle, be used to correct for the effect or at least find the 
sign of the effect.

\subsection{Anisotropies in the power spectrum} 

The light cone effect introduces an anisotropy in the full 3D 21
  cm power spectrum. For a fixed $k$-mode, the power spectrum will
  depend on $k_{\parallel}$, the LOS component of $k$. Peculiar
  velocities and the Alcock-Paczynski effect are the other major
  sources of anisotropies in the 21 cm power spectrum. In order to
  understand the anisotropic power spectrum and to separate the
  physics from astrophysics \citep{barkana05} each effect should be studied in
  detail. Though the first generation of low frequency radio
  telescopes (i.g, LOFAR, GMRT, MWA ) are unlikely to able to measure
  the anisotropies in the 21 cm power spectrum, this will be the
  ultimate goal of such measurements. 

  Fig. \ref{fig:ps-aniso} plots the ratio $\Delta^2_{\rm
    3Dlc}/\Delta^2_{\rm 3Dcc}$ as a function of $\mu^2$ for different
  $k$-modes for two central redshifts $z_{\rm c}=8.76$ (left panel)
  and $10.02$ (right panel) for L3 simulations.  Here
  $\mu=k_{\parallel}/k$. In the left panel ($z_{\rm c}=8.76$) we see
  that the ratio $\Delta^2_{\rm 3Dlc}/\Delta^2_{\rm 3Dcc}$ increases
  from $\sim 0.7 $ (at $\mu^2=0.1$) to $1.1$ (at $\mu^2=0.9$) for
  $k=0.16 \, {\rm Mpc^{-1}}$. For $k=0.3 \, {\rm Mpc^{-1}}$, the ratio
  increases from $\sim 0.8$ to $1$ for the same $\mu^2$ range. For
  higher $k$-modes the degree of anisotropy decreases and the power
  spectrum is becoming more isotropic. We do not see any significant
  anisotropies for the central redshift $z_{\rm c}=10.02$ (right
  panel) where the neutral fraction $x_{\rm HIc}=0.86$. The other redshifts
  of the L3 simulation also do not show significant anisotropies and the
  results for the L1 simulation are similar to L3. 

  We do not try to quantify the anisotropies further as we see the
  curves are not very smooth due to the small number of modes at large
  $k$.  Our results are sample variance limited and should be
  considered as qualitative rather than quantitative. Larger
  simulation volumes are needed to quantify the anisotropies more
  precisely.  \citet{barkana06} (fig. 2) reported significant
  anisotropies at large scales ($r=100 \, {\rm Mpc}$) for neutral
  fraction $x_{\rm HI}=0.25$ in their Pop III reionization
  model. This is qualitatively consistent with our results.

\subsection{1D LOS power spectrum}
 Recent results by \cite{harker10} show that the dimensionless 1D LOS power spectrum $\Delta^2_{\rm{1D,LOS}}(k)$ \footnote{$\Delta^2_{\rm{1D,LOS}}(k)= k P_{\rm{1D,LOS}}(k)/\pi$ \citep{peacock}} can be extracted more accurately as there is no large scale bias (which may arise due to foreground subtraction) and smaller error bars in the recovered 1D LOS power spectrum \citep[shown in figure 11 of][]{harker10}. The 1D LOS power spectrum can also extend to smaller scales because of the higher resolution in the frequency direction. Motivated by this, we also study the effect of evolution on the 1D LOS power spectrum.  Figure \ref{fig:ps1de_f10} shows the effect in the 1D LOS power spectrum for the L1 case. Interestingly, the 1D LOS power spectrum is hardly affected by light cone effects, except near the end of reionization. We do not show results for the L3 simulation as these are very similar to L1. This result adds one more advantage to the strategy of measuring 1D LOS power spectrum.  However, the 1D LOS power spectrum is very flat on large scales due to the aliasing of high $k$-modes. Therefore, although it can be more accurately measured, it may be difficult to see the impact of ionized bubbles and extract the reionization physics. 

We would like to mention here that besides the simulations L1, L3 we also analysed the L2 simulation \citep[for more details see][]{iliev11} where the reionization is much more gradual and overlap happens at a redshift $z=6.5$. The evolution is thus relatively slow and we see that the light cone effect is smaller in amplitude but otherwise has similar features as what we presented above for the cases L1 and L3.

\subsection{Comparison to previous work}

 \citet{barkana06} analytically studied the light cone effect
  using the two point correlation function $\xi(r, \mu)$, rather than
  considering the power spectrum. Since they considered different
  reionizations models, redshift range and the cosmological
  parameters, as well as another diagnostic, we can here only provide
  a qualitative comparison.

  \citet{barkana06} limited their investigation to the late stages of
  reionization ($x_{\rm HI}<0.5$) and find that the effect is
  significant from the time when the reionization is $\sim 70\%$
  complete to its end (see their fig. 4) and that large scales are
  affected more than small scales. We find similar results as we see
  the largest differences in the power spectra at large scales for the
  later stages of reionization. We also find substantial differences
  ($10-30\%$) in the first half of reionization, but this phase was
  not considered in \citet{barkana06}.

  For a fixed correlation length $r$ \citet{barkana06} showed that the
  correlation function $\xi(r, \mu)$ for $\mu=1$ is identical to the
  value for $\mu=0$ around $x_{\rm HI}=0.5$, suppressed close to the
  end of reionization, and enhanced in the intermediate period. As
  explained above, $\mu=1$ corresponds to the LOS direction and thus
  measures the light cone effect. Therefore, for a fixed length scale
  $r$ \citet{barkana06} found the light cone effect to have a negligible
  impact before and around $x_{\rm HI}=0.5$, to suppress the correlation
  function at the end of reionization but to enhance it in the
  intermediate period. This is exactly what we find. In the L3
  simulations, we find that the power spectra are suppressed during
  the late stages of reionization but enhanced before that.

\section{Toy models}

To understand the results from the previous section we consider here two analytical toy models of reionization.

\subsection{Toy model 1}

In this toy model we consider a very simple scenario. We consider $N$ number of spherical, non-overlapping and randomly placed ionized bubbles in a uniform H I medium in the coeval cube.  The spherically averaged 3D power spectrum for such a scenario can be written as
\be
P_{\rm {3D}}(k)=\sum_{i=1}^N V_i^2 W^2(kR_i)
\e
where $V_i=\frac{4}{3} \pi R_i^3$ and $W(kR)$ is the spherical top hat window function defined as
\be
W(kR)=\frac{3}{k^3R^3}[\sin(kR)-kR\cos(kR)]
\e
Now $P_{\rm{3D}}(k)=\sum_{i=1}^N V_i^2$ for $k<1/R_{\rm{max}}$, where $R_{\rm{max}}$ is the radius of the biggest bubble in the cube since $ W(x) \approx 1$ for $x<1$.

For the coeval cube we assume that all bubbles are of the same size $V_o$, therefore the power spectrum can simply be written as,
\be
P_{\rm {3Dcc}}(k)=N V_o^2.
\e
Because of the evolution effect in the light cone cube, bubbles at  the back side will appear smaller and bubbles at the front side will appear  bigger and in addition their shapes could be somewhat elongated along the LOS \citep[see Figure 1 of][]{majumdar10}. To make our calculations simpler we assume the bubbles in the light cone cube are spherical but have different sizes $V_o+\Delta V_i$. As we saw in the simulation results, the global ionization fraction for light cone cubes is almost the same as in the coeval cube at the central redshift, so we assume $\sum_{i=1}^N \Delta V_i=0$. The spherically averaged 3D power spectrum for the light cone cube at larger scales is then given as,
\begin{eqnarray}
P_{\rm {3Dlc}}(k)&=&\sum_{i=1}^N (V_o+ \Delta V_i)^2  \nonumber \\
&=& P_{\rm {3Dcc}}(k)+2V_o \sum_{i=1}^N \Delta V_i+ \sum_{i=1}^N\Delta V_i^2 \nonumber \\
&=& P_{\rm {3Dcc}}(k)+\sum_{i=1}^N\Delta V_i^2
\end{eqnarray}
The above equation explains two major features we see in the simulation. First it explains why the light cone effect is relatively small. We see that the effect cancels out in the linear order. Only the 2nd order term $\sum_{i=1}^N\Delta V_i^2$ survives the averaging and affects the light cone power spectrum. So in this sense the light cone effect is a `2nd order effect' in the spherically averaged power spectrum. Second, because $\sum_{i=1}^N\Delta V_i^2$ is always positive the power is always enhanced at larger scales which is exactly what we see in the simulation.  When the bubble sizes are not identical in the coeval cube, we can subdivide the entire range of bubble sizes into small size bins. The above analysis would then be applicable to each individual size bin and thus to the entire range.

Our second toy model considers a slightly more realistic but still quite simple reionization scenario.

\subsection{Toy model 2}

In this toy model, we consider a reionization model in which spherical ionized bubbles of different sizes are placed randomly. If there is no overlap between ionized bubbles then the ionized  fraction would be,
\be
Q=\int dR \frac{dn}{dR} V(R)
\label{eq:ionfrac}
\e 
where $n(R)$ is the number density of bubbles of size $R$. But in practice randomly placed bubbles will overlap with each other and expand further  to conserve the emitted photon numbers. We neglect the fact of further expansion of bubbles and so the actual  ionized fraction which would be less than the above can be calculated using \citep{fur04}
\be
{\bar x_i}=1-\exp(-Q).
\e
The spherically averaged 3D power spectrum can be calculated from the two point correlation function using the relation
\be
P_{\rm {3D}}(k)=4 \pi \int_{r=0}^{\infty}dr r^2 \xi(r) \frac{\sin(kr)}{kr}
\e
Here we would like to mention that the evolution makes the correlation function  $\xi$ anisotropic i.e, a function of both $r$ and $\mu$ \citep{barkana06,sethi08}. Since our aim is to study the light cone effect on the spherically averaged power spectrum and qualitatively understand the main features we see in the simulation results, we assume $\xi$ to be isotropic. We expect this approximation not to affect our conclusions. The correlation function $\xi(r)$ can be decomposed into \citep{zal04}
\be
\xi(r)= \xi_{xx}(1+\xi_{\delta \delta})+\bar{x}_{H I}^2\xi_{\delta \delta}+\xi_{x \delta}(2 \bar{x}_{H I}+\xi_{x \delta})
\e
where $\xi_{xx}$, $\xi_{\delta \delta}$ and $\xi_{x \delta}$ are the correlation functions of ionization field, density field and the cross-correlation between two fields respectively. 

\begin{figure}
\includegraphics[width=.33\textwidth, angle=270]{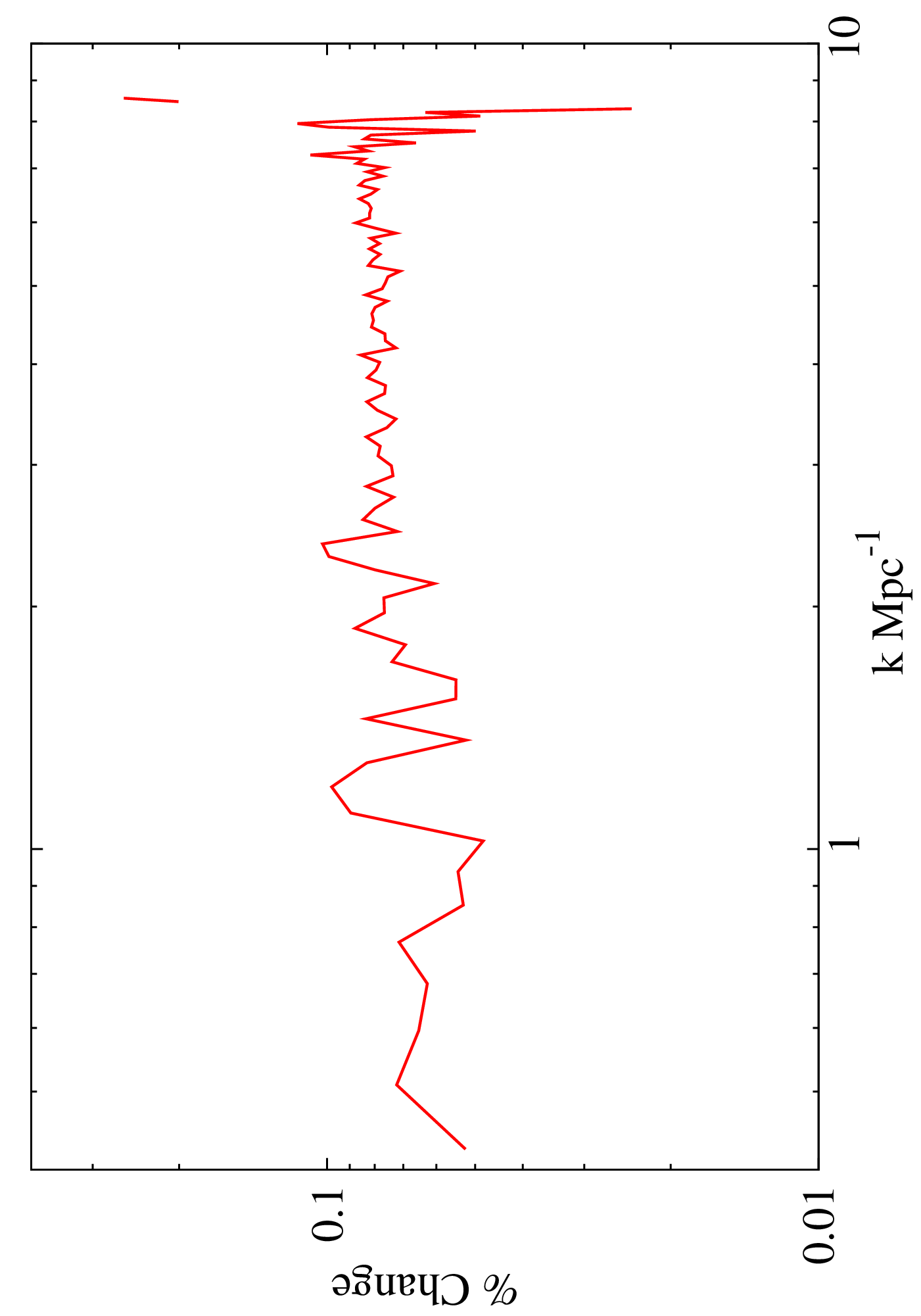}
\caption{The relative change in the 3D 21-cm power spectrum $\Delta^2_{\rm{3D}}(k)$ due to the combined effect of linear growth of structure and adiabatic expansion of the Universe. The ordinate is the quantity $100\times(\Delta^2_\mathrm{3Dw}-\Delta^2_\mathrm{3Dwo})/\Delta^2_\mathrm{3Dwo}$, where `w/wo' means the effect is/is not included. We use very high redshift simulation cubes (before the 
reionization starts) to calculate this.}
\label{fig:linearpower}
\end{figure}

In the density field correlation function $\xi_{\delta \delta}$, two
quantities change with redshift: 1) density fluctuations grow with
time, 2) the mean density decreases because of the
expansion of the Universe. Thus the evolving $\xi_{\delta \delta}$
in principle would contribute to the light cone effect. In the linear
regime the two quantities together scale as $\sim (1+z)^{-0.5}$. For a
distance of $163 \, \rm{cMpc}$ they jointly change $\sim 3
\%$ along the LOS.  We use very high redshift simulation cubes (before the 
reionization starts) and find $\lesssim 0.1 \%$ enhancement in the 3D
power spectrum on almost all scales (Figure
\ref{fig:linearpower}). This result agrees with \citet{mcquinn06} who
predicted a constant enhancement in the power spectrum (see their
Appendix A). The contribution of the evolving $\xi_{\delta \delta}$ to
the total light cone effect on the power spectrum is therefore
negligible and hence we ignore this term in the rest of our analysis.

The evolution of $\xi$ is thus mainly dominated by the $\xi_{xx}$ on scales larger or comparable to the size of ionized bubbles. For the rest of our analysis we only consider the term $\xi_{xx}$. The function which is defined as $\xi_{xx}(r)=\langle x_1x_2\rangle-\bar{x}_i^2$ should be zero both for $\bar{x}_i=0$ and $\bar{x}_i=1$. It also should satisfy the boundary conditions \citep[for details see][]{zal04}

\[\xi_{xx}(r) = \left\{ 
\begin{array}{l l}
  \bar{x}_i-\bar{x}_i^2 & \quad \mbox{for $r \rightarrow 0$}\\
  0 & \quad \mbox{for $r \rightarrow \infty. $}\\ \end{array} \right. \]

The correlation function can be calculated  for a given bubble distribution as 
\citep{fur04}
\begin{eqnarray}
\lefteqn{\langle x_1 x_2\rangle(r) = } \nonumber \\
&& \left( 1-\exp \left[ - \int dR \frac{dn}{dR} V_o (R) \right] \right) \nonumber \\
&& {} +\exp \left[ - \int dR \frac{dn}{dR} V_o (R) \right] \nonumber \\
&& {} \times \left( 1-\exp \left[ - \int dR \frac{dn}{dR} [V(R)-V_o (R)] \right] \right)^2 
\end{eqnarray}
where $V_o(R,r)$ is the volume of the overlap region between two ionized regions centered a distance $r$ apart. The function can be written as
\[V_o(R,r)= \left\{ 
\begin{array}{l l}
4 \pi R^3/3- \pi r[R^2-r^2/12] & \quad \mbox{for $ r<2 R$}\\
0 & \quad \mbox{for $r>2 R$} \\ \end{array} \right. \]

In the next subsection we present some bubble size distributions measured from simulation. We will then model the bubble distribution and use that for the subsequent analysis. 

\begin{figure}
\includegraphics[width=.33\textwidth, angle=270]{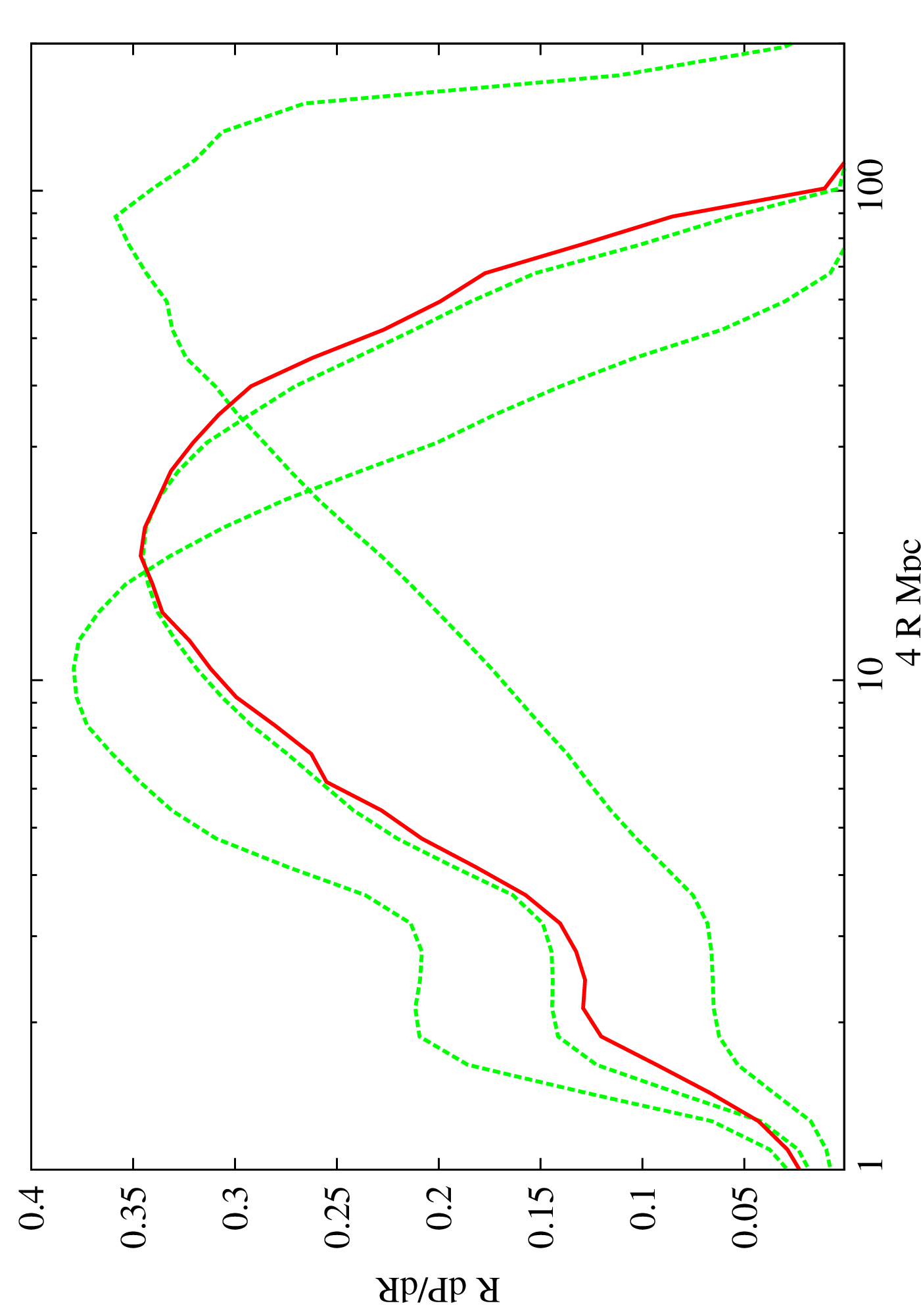}
\caption{Bubble size distribution calculated from the simulated cubes around redshift $z=9.09$ for L3 model. The dashed lines show the bubble distributions for three coeval cubes at redshifts $z=9.38, 9.09$ and $8.76$ (from left to right) corresponding to the back, middle and the front side of the light cone cube centered around redshift $9.09$. The solid line (red) show the distribution for this light cone cube.}
\label{fig:bub-dis-sim}
\end{figure}

\begin{figure}
\includegraphics[width=.33\textwidth, angle=270]{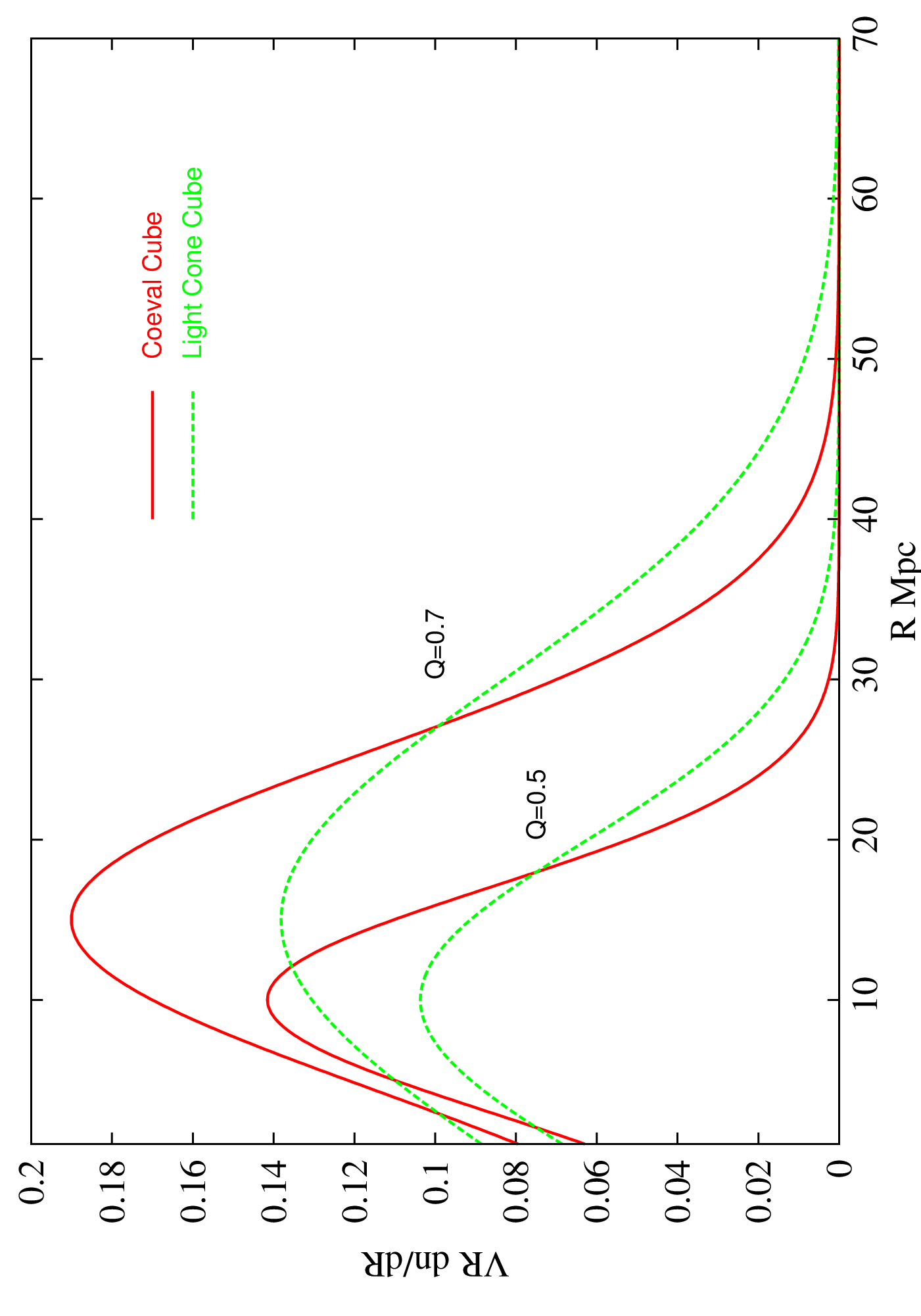}
\caption{Bubble size distributions used to calculate the power spectrum in the Toy model 2. Solid (red) and dashed (green) lines represent  coeval and light cone cubes respectively.}
\label{fig:bub-dis-toy2}
\end{figure}

\begin{figure}
\includegraphics[width=.33\textwidth, angle=270]{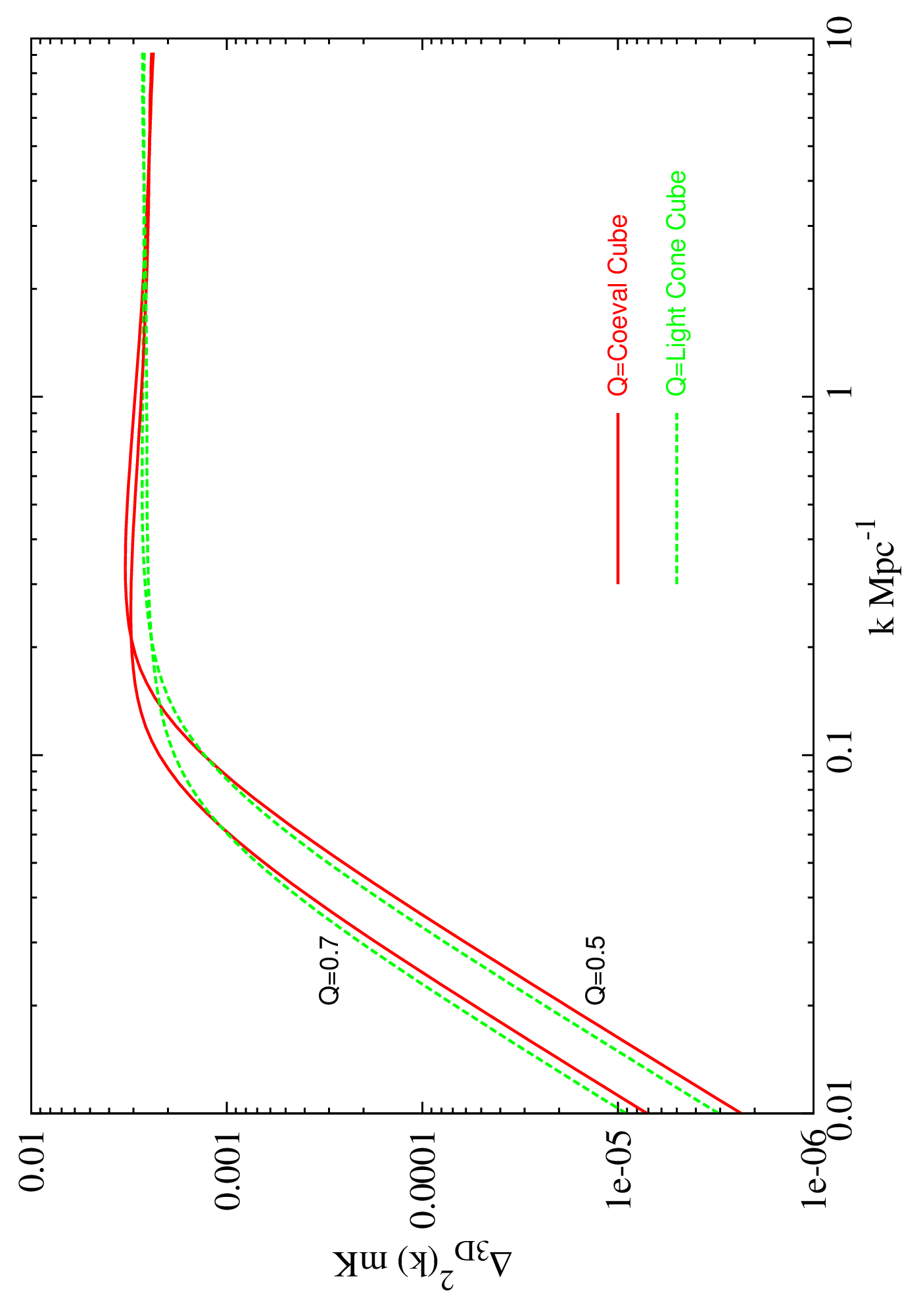}
\caption{The power spectra for the different bubble size distributions from Figure \ref{fig:bub-dis-toy2}. The y-axis is scaled arbitrarily.}
\label{fig:Pk_toy2}
\end{figure}

\subsubsection{Bubble size distribution and its evolution}

We calculate the bubble size distribution from the simulation using the spherical average method \citep[see][ for more details on different bubble size estimates]{friedrich11}.
Fig.~\ref{fig:bub-dis-sim} shows the bubble size distribution $R \, dP/dR$\footnote{This is quantity is essentially same as $V(R)(dn/dR)$ and follows the condition $\bar{x}_i=\int (dP/dR)\,dR$} for coeval cubes at three redshifts corresponding to the back, middle and front side of a light cone cube centered around redshift 9.09. It also shows the bubble size distribution in the light cone cube centered around 9.09. The coeval distributions at the three redshifts differ considerably. For example the radii at which the bubble distribution peaks are 10, 20, and 90 cMpc. Interestingly the bubble size distribution for the light cone cube is very similar to the coeval box at the central redshift. In the light cone cube the bubble size distribution would be the average of those of the coeval cubes in the redshift range $z_c \pm \Delta z$ where $2 \Delta z$ is the extent of the cube along redshift axis. Although the bubbles in the light cone cube are smaller/larger in the back/front side compared to the coeval cube, the average bubble distribution in the whole light cone cube is very similar to the coeval cube of the central redshift. Because of this `averaging effect' the light cone effect is small even though there is a substantial evolution in the bubble distribution across the box. 

We investigate further and to make the following calculations simpler we parameterize $V(R)\frac{dn}{dlnR}=A \exp(-(R-R_c)^2/\sigma_R^2)$. This is motivated by Fig.~\ref{fig:bub-dis-sim} \citep[see also Figure 2 in][]{fur04}. 
We normalize the function using Eq.~\ref{eq:ionfrac}.

Now consider the case around the central redshift $z_c$.  Since the light cone cube covers the redshift range $z_c \pm \Delta z$ it will have slightly more bubbles both at the large and small bubble size ends than the coeval cube at redshift $z_c$. This is exactly what we see in the simulation (Figs.~\ref{fig:image} and \ref{fig:bub-dis-sim}).  

As we mentioned in Section 4.1 the average neutral fraction $x_{\rm{H I}}$ for a coeval cube at redshift $z_c$ and for a light cone cube centered around redshift $z_c$ is almost the same. We consider reionization for two values of $Q=0.5$ and $0.7$ (see Eq.~\ref{eq:ionfrac}). Parameters for the bubble distribution are summarized in Table 3. Figure \ref{fig:bub-dis-toy2} shows the bubble distribution for  $Q=0.5$ (solid) and $0.7$ (dashed) for the coeval cube and the light cone cube. As we discussed above, there will be more large and small size bubbles in the light cone cube compared to the coeval cube, we approximate this by increasing $\sigma_R$ for the light cone cube. The bubble size at which the quantity $V\frac{dn}{dlnR}$ peaks has been kept same for both for a fixed  $Q$. We also see in the Fig.~\ref{fig:bub-dis-toy2} that for $Q=0.5$ the number density of bubbles of size $R_b>18 \, \rm{cMpc}$ is higher in the light cone cube than the coeval cube. The `cross over radius' i.e, the bubble radius beyond which the number density becomes higher than in the coeval cube is $18 \, \rm{cMpc}$. For  $Q=0.7$ the cross over radius ($\sim 27 \, \rm{cMpc}$) is higher than for $Q=0.5$. This is because for higher  $Q$ the characteristic bubble size increases. Obviously, the exact distribution  could be different but the general features such as the increase of characteristic bubble size and cross over radius for larger $Q$ values, and larger bubbles in the light cone  cube than the coeval cube, are likely to be true in all reionization scenarios where stars/QSOs are dominant sources. Since our aim is to qualitatively understand the effect of evolution, we can use these
simplified distributions.


\begin{table}
\caption{Parameters for the bubble size distributions in Toy model 2.}
\label{tab:table_bub_dis}
\begin{tabular}{|c|c|c|c|}
\hline
Box & $Q$ & $R_c$ (cMpc) & $\sigma_R$ (cMpc) \\ 
\hline
coeval cube& 0.5& $10$ & $10$ \\
\hline
Light cone cube &  0.5 & $10$ & $14$ \\
\hline
coeval cube & 0.7 &  $15$ & $15$ \\
\hline
Light cone cube &  0.7 & $15$ & $21$ \\
\hline
\end{tabular}
\end{table}

Figure \ref{fig:Pk_toy2} plots the power spectrum for the bubble distribution models we describe above. We see that there is a scale $k_{\rm{cross-over}}$ below ($k<k_{\rm{cross-over}}$) which the light cone cube has more power than the coeval cube and above ($k>k_{\rm{cross-over}}$) which it is the other way around. This is because the number of bubbles in the light cone cube  below the cross-over radius is less than in the coeval cube. We call this scale the `cross-over mode'. We see this feature in Fig.~\ref{fig:Pk_toy2} where the cross-over modes are $k_{\rm{cross-over}}=0.097 \, \rm{Mpc}^{-1}$ and $0.063 \, \rm{Mpc}^{-1}$ for $Q=0.5$ and $0.7$ respectively. The cross-over mode is thus seen to shift towards larger scales (smaller $k$) as $Q$ increases. This is because the cross-over radius is larger for $Q=0.7$. Based on these results we find the empirical relation between the cross-over radius and the cross-over mode to be
\be
R_{\rm{cross-over}}=\frac{1.7}{k_{\rm{cross-over}}}
\e

This simple toy model explains the two main features seen in the simulation results
\begin{enumerate}
\item The power spectrum in the light cone cube is enhanced/suppressed on large/small scales compared to the one from the coeval cube at the central redshift.
\item The cross-over mode shifts towards large scales as reionization proceeds.
\end{enumerate}
As we pointed out in Sect.~\ref{sect:sim_results}, we do not see a large effect when the mean ionization fraction is around $50 \%$, even though the evolution is across the light cone cube is substantial at that stage. We can now understand this to be because the cross-over mode shifts towards larger scales as reionization proceeds and around $50 \%$ ionization the cross over scale is already almost the same as lowest mode that we can measure from our simulation volume, even though it has a size of 163~cMpc. We therefore predict that for a larger simulation volume, enhanced power on scales $k < 0.08$~Mpc$^{-1}$ will be found.

\begin{figure*}
\begin{center}
\includegraphics[width=.46\textwidth, angle=270]{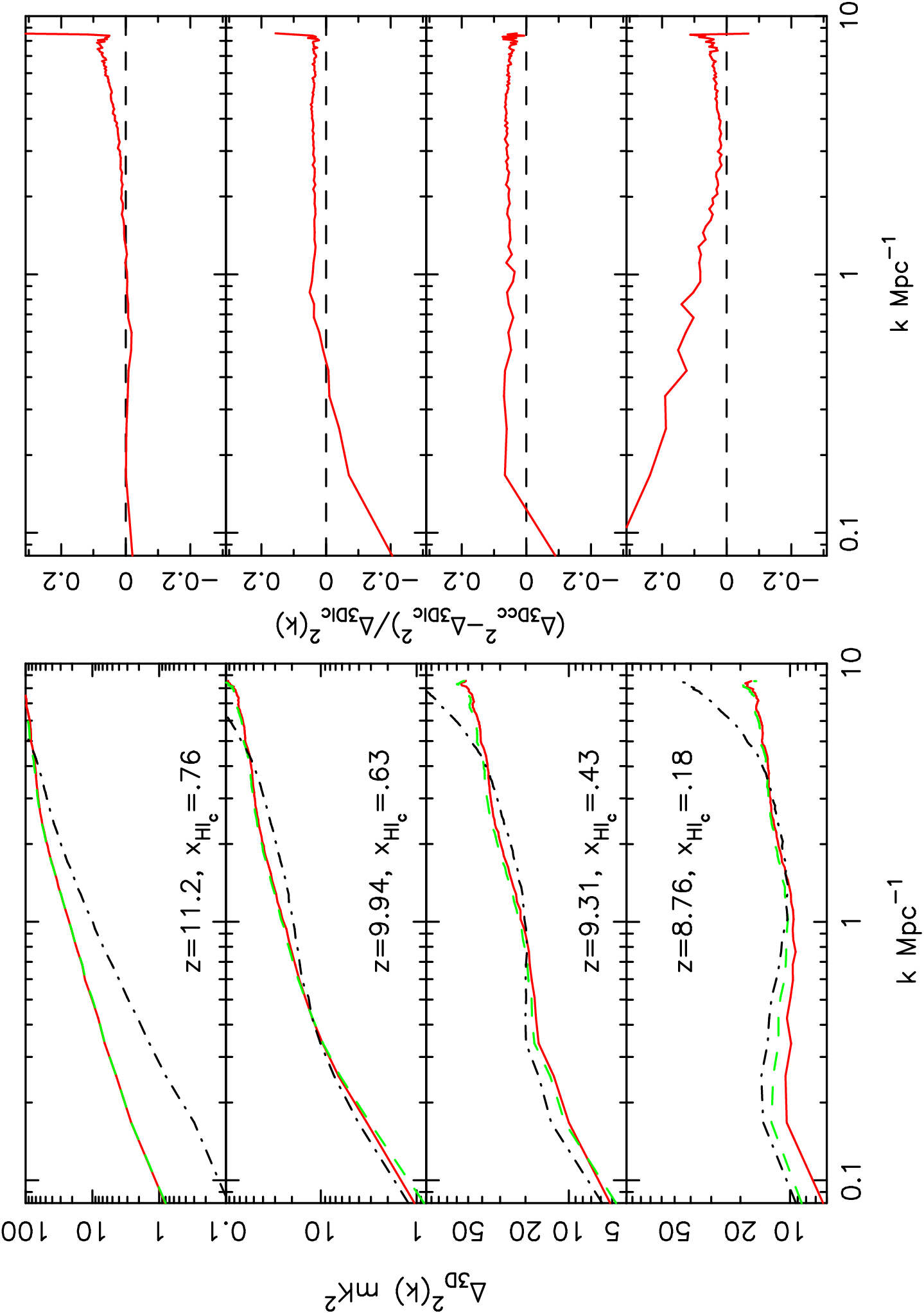}
\caption{The effect of peculiar velocity (pv) on the light cone (lc) effect in the L1 case at different redshifts. Left panels: Dot dashed, dashed and solid lines show the power spectrum for cubes without pv but with lc, with pv but without lc and with both pv and lc, respectively. The right panels plot $(\Delta^2_{\rm{3Dcc}}-\Delta^2_{\rm{3Dlc}})/\Delta^2_{\rm{3Dlc}}$ where we incorporated the peculiar velocity in both $\Delta^2_{\rm{3Dcc}}$ and $\Delta^2_{\rm{3Dlc}}$.} 
\label{fig:ps3de_f10-pec-v}
\end{center}
\end{figure*}

\subsection{Taylor expansion of the power spectrum evolution}
In addition to the more heuristic models given above, the following approach also helps in understanding some of the trends we see in Figure \ref{fig:ps3de_f10} and \ref{fig:ps3de_f25}. We find that the coeval cube power spectrum $\Delta^2_{\rm 3Dcc}(k,z)$ for a given mode ${\rm k}$ changes very smoothly with redshift $z$. So we expand $\Delta^2_{\rm 3Dcc}(k,z)$ in a Taylor series around a central redshift $z_c$ as
\be
\Delta^2_{\rm 3Dcc}(k,z)=\Delta^2_{\rm 3Dcc}(k,z_c)+a(\Delta L)+b(\Delta L)^2+c(\Delta L)^3+..,
\label{eq:pk-taylor}
\e
where $\Delta L$ is the comoving distance from the central redshift $z_c$ to redshift $z$ and the parameters $a=\left (\frac{d \Delta^2_{\rm 3Dcc}}{dL}\right)_{z_c}$, $b=\frac{1}{2!}\left (\frac{d^2 \Delta^2_{\rm 3Dcc}}{dL^2}\right)_{z_c}$, $c=\frac{1}{3!}\left (\frac{d^3 \Delta^2_{\rm 3Dcc}}{dL^3}\right)_{z_c}$. We ignore the higher order terms. Next we calculate the light cone power spectrum $\Delta^2_{\rm 3Dlc}(k,z_c)$ by taking average of $\Delta^2_{\rm 3Dcc}(k,z)$ in the range $\pm L/2$ around redshift $z_c$ using
\be
\Delta^2_{\rm 3Dlc}(k,z)=\frac{1}{L} \int^{L/2}_{-L/2}\Delta^2_{\rm 3Dcc}(k,z) dL,
\label{eq:pk-lc} 
\e 
where $L$ is the comoving LOS width. The above equation can be simplified to
\be
\Delta^2_{\rm 3Dlc}(k,z)=\Delta^2_{\rm 3Dcc}(k,z_c)+b\frac{ L^2}{12}.
\label{eq:pk-lc1}
\e 
We see that the linear term ($a \Delta L$) and all terms with odd powers cancel out and only the quadratic term ($b (\Delta L)^2$) and the other terms with even powers survive the averaging process. This supports our argument that the light cone effect is a `2nd order effect' and that linear trends in the evolution of
the power spectrum average out. The fractional change in the power spectrum due to the light cone effect is given by $\frac{bL^2}{12\Delta^2_{\rm 3Dlc}}$. Positive/negative values of the parameter $b$ denote that light cone power is suppressed/enhanced compared to the coeval value.

To test this quadratic approximation we use simulation L3 and fit the
polynomial Eq.~\ref{eq:pk-taylor} for a given mode $k$ around three
different central redshifts, taking $L=163 \, {\rm cMpc}$. Using the
values of $b$ we calculate the percentage change in the power
spectrum and also measure the actual percentages from the simulation
results. Values for three different $k$ modes are given in
Table~\ref{tab:taylor}. From these it can be seen that the quadratic
expansion correctly predicts the sign of the parameter $b$ and
reproduces the trends seen in the simulations. During the early phases
the match with the simulations is quite good, but at later stages it
under predicts the changes. Most likely the discrepancies are due to
the neglect of the higher order terms.
\begin{table}
\caption{Comparison between the quadratic expansion and simulation results.
Listed are the relative sizes of the light cone effect as predicted by
the quadratic expansion (Q) and measured in the simulation (S).}
\label{tab:taylor}
\begin{tabular}{|c|c|c|c|c|c|c|}
\hline
$k$ (Mpc$^{-1}$) &\multicolumn{2}{|c|}{0.081} & \multicolumn{2}{|c|}{0.167} & \multicolumn{2}{|c|}{1.02}\\ 
\hline
$z_c$ & Q & S & Q & S & Q & S \\
\hline
 $8.76$ & 21\% & 52\% & 16\% & 24\% & 4\% & 12\% \\
 $9.31$ & -3.3\% & -9\% & 2.5\% & 6\% & 2.4\% & 3\% \\
 $10.02$ & -24\% & -30\% & -11\% & -10\% & 0.6\% & 1\% \\
\hline
\end{tabular}
\end{table}

\section{Effect of peculiar velocity on the light cone effect}
\label{pecvels}
The peculiar velocity of the IGM gas influences the 21-cm power spectrum \citep{bharadwaj01, bharadwaj04}. During the dark ages when the H I density is expected to trace the DM density, the spherical averaged power spectrum is enhanced by a factor of 1.87 at linear scales. As reionization proceeds, the relative contribution of the peculiar velocity to the 21-cm power spectrum  changes considerably with redshift. For inside-out reionization scenario the peculiar velocity could increase the 21-cm power spectrum by a factor of $\sim 5$ \citep[see][ Fig. 3]{mao11} during a short period in the beginning of reionization \citep[$x_i <0.2$, see][ Fig. 3]{mao11}. When reionization is at its $\sim$50\% phase, peculiar velocity effects slightly decreases the 21-cm power spectrum on the large scales relevant for the first generation of EOR experiments. In other words, peculiar velocity effects change the evolution of 21-cm power spectrum and hence could affect the light cone effect. We briefly investigate this here.

The method for taking the peculiar velocity into account when constructing the light cone cube was outlined in \cite{mellema06} and described in detail in \citet{mao11}; in the terminology of the latter we use the MM-RMM(1$\times$RT) scheme. The left panel of Figure \ref{fig:ps3de_f10-pec-v} shows the 21-cm power spectrum with peculiar velocity for coeval cubes at three different redshifts (center and two ends) as well as for the light cone cube. The right panel shows the relative difference $ (\Delta_{\rm{3Dcc}}^2-\Delta_{\rm{3Dlc}}^2)/\Delta_{\rm{3Dlc}}^2$. The figure looks mostly very similar to Figure \ref{fig:ps3de_f10} where we did not include any peculiar velocity effects, the exception being the earliest stages, at redshift $z=11.20$. Here the case with peculiar velocity shows a negligible light cone effect whereas the case for no peculiar velocity shows a $\sim 10 \%$ difference in the power spectrum at large scales. 
The reason for this is that in the beginning of reionization bubble growth is relatively slow and evolution is dominated by the peculiar velocity. As reionization proceeds the evolution is mainly dominated by the growth of ionized bubbles and hence the peculiar velocity has almost no impact on the light cone effect. Note that this does not mean that the inclusion of peculiar velocity does not affect the power spectrum. In fact peculiar velocity causes comparable or even larger changes than the light cone effect. A more thorough exploration of the effects of peculiar velocity will be presented in future work.

\section{Conclusions and discussion}
We investigate the effect of evolution on the 
3D and 1D LOS 21-cm power spectra during the entire period of reionization. We use three different EoR simulations in a volume of 163 cMpc on each side, one in which reionization is more gradual, ending at $z\approx 6.5$ and two in which it is more rapid, ending at $z\approx 8.5$. In one of these rapid simulations, reionization is driven by more massive sources, leading to relatively larger ionized bubbles. Below we summarize our results:
\begin{itemize}
\item For the cases we studied, the spherically averaged power
  spectrum changes up to $\sim 50\%$ in the $k$ range $0.08$ to $9 \,
  \rm{Mpc}^{-1}$ using a redshift interval corresponding to the
    full extent of our simulation volume, $(163 \, {\rm cMpc})$. As
    expected, for smaller redshift bins the effect is found to be
    smaller. Large scales are affected more and the effects
  at smaller scales are minor.

\item Substantial evolution of the mean mass averaged neutral fraction $x_{\rm{H I}}$, rms variations in the 21-cm signal, and bubble size distribution along the LOS axis are averaged out in the spherically averaged power spectrum. This averaging effect makes the light cone effect relatively small compared to the evolutionary changes along the LOS axis.

\item  We can detect anisotropies in the the full 3D power spectra
    on large scales in the later stages of reionization, but are unable
    to quantify the $\mu$-dependence of this effect with the
    simulations available to us.

\item The bubble size distribution in the light cone cube centered around redshifts $z_c$ is remarkably similar to the bubble size distribution in the coeval cube at $z_c$, even if there is substantial evolution in the ionized fractions along the LOS. This is the reason why we see a relatively small effect on the 21-cm power spectrum compared to the amount of change in the $x_{\rm{H I}}$ and the rms of the 21-cm signal.

\item The large scale power is enhanced and the small scale power is suppressed most of the time except at the final phase of reionization where the power spectrum is suppressed at all scales we can measure in our simulations. In other words, there is a `cross-over mode' $k_{\rm{cross-over}}$ below and above which the power is enhanced and suppressed respectively. The  cross-over mode $k_{\rm{cross-over}}$ shifts towards lower $k$-mode (large scale) as reionization proceeds.

\item Surprisingly we see very little effect when reionization is $\sim 50 \%$ complete and there is a rapid evolution in the $x_\mathrm{H I}$ and the rms. We argue that at this stage of reionization the cross-over mode $k_{\rm{cross-over}}$ is already comparable to the lowest $k$ mode we can measure from the simulation and enhancement of power should be present at larger scales than that.

\item Despite the fact that the reionization histories differ considerably between the three simulations, we see quite similar results.

\item Growth of structures with redshift and the expanding background enhance the power spectrum by  $\sim 0.1 \%$ for our $163$ cMpc cube. Its evolution is 
therefore dominated by the ionization field during the reionization. 

\item An analytical toy model (Toy model 1) can explain the large scale power enhancement due to light cone effect as well as its smallness.

\item A second analytical toy model (Toy model 2) for a light cone cube with more large bubbles beyond some cross-over radius $R_{\rm{cross-over}}$ and less bubbles below that, can explain all the features we see in the simulation results. 

\item The presence of more large bubbles and fewer small bubbles of size $<R_{\rm{cross-over}}$ is responsible for the enhanced/suppressed power on scales below/above $k_{\rm{cross-over}}$. The fact that the cross-over scale shifts towards lower $k$ as reionization proceeds is because the  cross-over bubble size $R_{\rm{cross-over}}$ increases as reionization proceeds. 

\item Interestingly we find that the light cone effect is less prominent in the 1D LOS power spectra. 

 \end{itemize}

We should note that instruments such as LOFAR and MWA are expected to measure down to $k\sim 0.01$~Mpc$^{-1}$, scales larger than we were able to analyze here ($k_{\rm{min}}=0.08 \, \rm{Mpc}^{-1}$). From our results we expect enhanced power on those scales in the light cone cube. Especially when reionization is around $\sim 50 \%$ we expect more enhanced power on these larger scales. Reionization simulations of even larger cosmological volumes would be useful to better understand the effects at those scales. On the other hand, the aforementioned telescopes will not reach beyond $k\sim 1$~Mpc$^{-1}$ making the small scale light cone effects observationally less relevant. 

 The removal of the large foreground signals of the EoR 21cm
  signal is expected to affect the large scale LOS modes
  $k_{\parallel}$ significantly. Although details about which scales
  will be affected depend on the subtraction technique used, it is
  obvious that if $L$ is the comoving length over which the foreground
  subtraction is performed, modes with $k_{\parallel} \lesssim 2
  \pi/L$ cannot be extracted \citep{mcquinn06}. The
  equivalent bandwidth for the simulation boxes we consider is $\sim
  10\, {\rm MHz}$ and it is likely that foreground subtraction techniques
  will use considerably larger bandwidths \citep[see e.g,][]{chapman12}.
  The same authors also show that foreground residuals do not affect
  the extraction of the 3D spherically averaged power spectrum over
  bandwidths of 8~MHz. However, the effects of foregrounds remain clearly
  an issue which requires careful consideration when considering LOS effects
  in the 21cm signal.

  In our simulations the spherically averaged power spectra are based
  on equal numbers of modes in the LOS and transverse
  directions. However, most of the ongoing and upcoming surveys will
  not sample the full range in the spatial and frequency directions
  for many $k$-modes. This is due to the fact that they have better
  resolution in the frequency (LOS) direction than in the spatial
  directions. For example the LOFAR core has a maximum baseline which
  corresponds a maximum transverse mode ${k_{\perp}}_{\rm max} \sim 1
  \, {\rm Mpc^{-1}}$. The intrinsic frequency resolution of the array
  is better than 1~kHz, but likely the observed data will be stored
  with $\sim 10 \, {\rm KHz}$ frequency resolution, equivalent to LOS
  mode ${k_{\parallel}}_{\rm max} \sim 35 \, {\rm Mpc^{-1}}$.  When
  using this resolution to calculate the spherically averaged power
  spectra, the LOS modes $k_{\parallel} \lesssim k$ will contribute
  more compared to the transverse modes for $k > {k_{\perp}}_{\rm
    max}$. Since the light cone effect makes the power spectra
  anisotropic i.e, different power for different combintions of
  $(k_{\perp}, k_{\parallel})$ for a given $k$, the lack of small
  scale transverse modes $k_{\perp} \sim k$, in principle, would
  affect the power spectra measurements at those $k$-modes. However,
  as shown in Fig. \ref{fig:ps-aniso}, small scales are hardly
  anisotropic due to the light cone effect so we do not expect those
  modes to be affected much due to the incomplete sampling of small
  scale modes. In addition, small scales $k \gtrsim 0.6 \, {\rm
    Mpc^{-1}}$ are expected to be dominated by system noise and are
  unlikely to be measured.

Based on our results, we conclude that the light cone effect is important especially at scales where the first generation of low frequency instruments are sensitive. It can bias cosmological and astrophysical interpretations unless this effect is understood and incorporated properly.

\section*{Acknowledgment}
We would like to thank the anonymous referee for his constructive
comments which have helped to improve the paper. Discussions with
Saleem Zaroubi and other members of the LOFAR EoR Key Science Project
have been valuable for the work described in this paper.  KKD is
grateful for financial support from Swedish Research Council (VR)
through the Oscar Klein Centre (grant 2007-8709). The work of GM is
supported by the Swedish Research Council grant 2009-4088. The authors
acknowledge the Texas Advanced Computing Center (TACC) at The
University of Texas at Austin for providing HPC resources, under NSF
TeraGrid grants TG-AST0900005 and TG-080028N and TACC internal
allocation grant ``A-asoz'', as well as the Swedish National
Infrastructure for Computing (SNIC) resources at HPC2N (Ume\aa,
Sweden), which have contributed to the research results reported in
this paper.  This work was supported in part by NSF grants AST-0708176
and AST-1009799, NASA grants NNX07AH09G, NNG04G177G and NNX11AE09G,
and Chandra grant SAO TM8-9009X. ITI was supported by The Southeast
Physics Network (SEPNet) and the Science and Technology Facilities
Council grants ST/F002858/1 and ST/I000976/1. KA is supported in part
by Basic Science Research Program through the National Research
Foundation of Korea (NRF) funded by the Ministry of Education, Science
and Technology (MEST; 2009-0068141,2009-0076868).

\appendix

\section{Effect of inclusion of $\mathrm{\k}(\k_{\rm x}=0, \k_{\rm y}=0,\k_{\rm z})$ modes  on the power spectrum}
\label{sec:vis}

\begin{figure}
\includegraphics[width=.33\textwidth, angle=270]{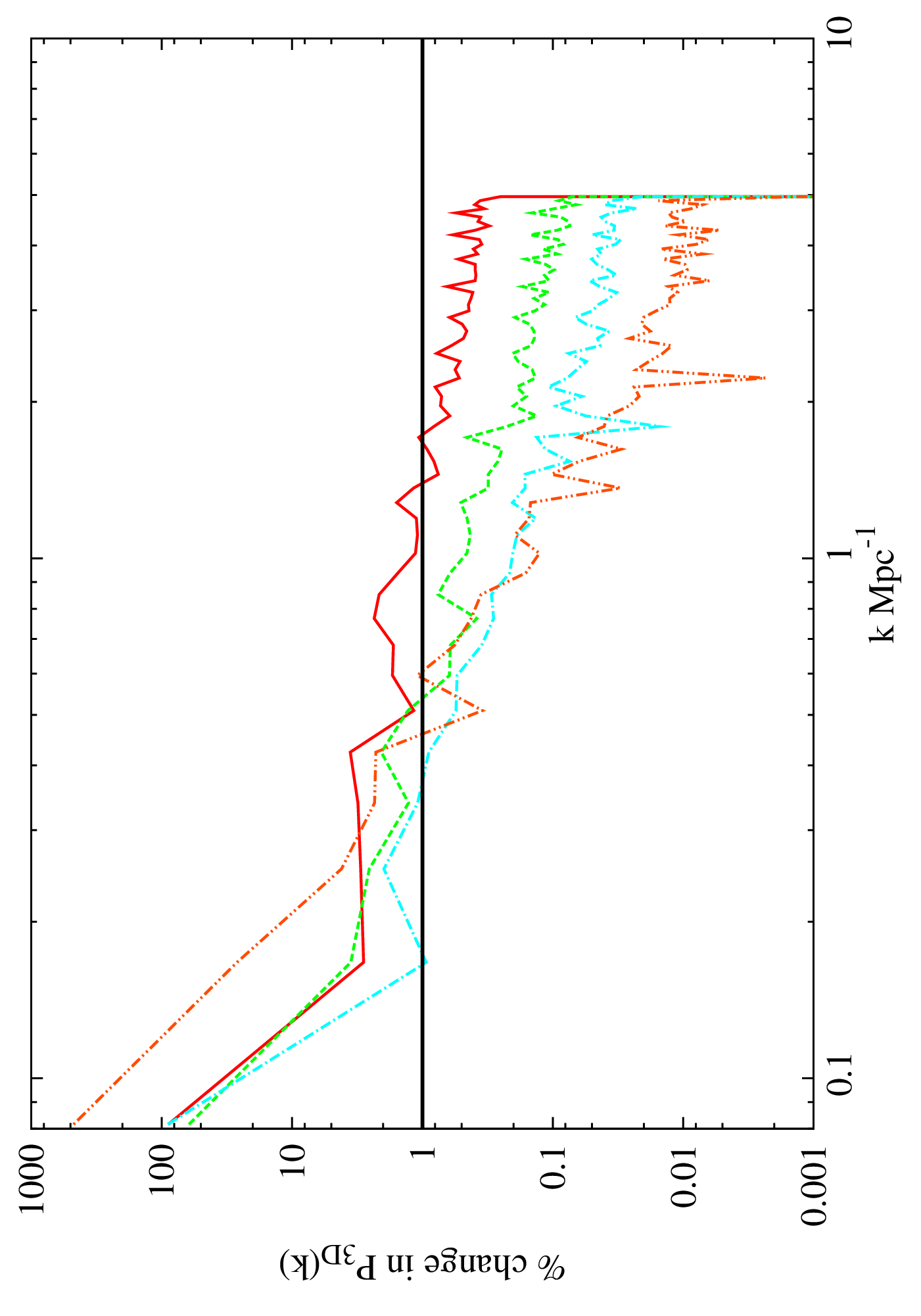}
\caption{Illustration of the effects of including $\mathrm{\bf {k}}(\k_{\rm x}=0, \k_{\rm y}=0,\k_{\rm z})$ modes  on the 3D power spectrum (simulation L1). The abscissa shows the quantity $100\times[\Delta^2_{\rm{3D-in}}-\Delta^2_{\rm{3D-ex}}]/\Delta^2_{\rm{3D-ex}}$, where `in/ex' stands for including/excluding the $k_x=k_y=0$ modes. Solid, dashed, dot-dashed and dot-dot-dashed lines are for redshifts 8.76, 9.31, 9.94 and 11.20 respectively. }
\label{fig:change_pk3dg}
\end{figure}

Radio interferometric experiments cannot measure the  modes at $k_x=k_y=0$ where $k_{(x,y)}=\frac{2 \pi (u,v)}{r}$, $u,v$ are two components of the baseline vector $\bf{U}$ and $r$ is the comoving distance. In order to predict the expected 21-cm power spectra for some reionization model or interpret the observed 21-cm power spectra the modes  $\mathrm{\bf {k}}(\k_{\rm x}=0, \k_{\rm y}=0,\k_{\rm z})$  should be excluded when the power spectrum is calculated from the simulated data. Figure \ref{fig:change_pk3dg} plots $100\times[P_{\rm{3D-in}}(k)-P_{\rm{3D-ex}}(k)]/P_{\rm{3D-ex}}(k)$ with $k$ for different redshifts for light cone cubes. Here $ P_{\rm{3D-in}}(k)$ and $P_{\rm{3D-ex}}(k)$ are the 3D power spectra including  and excluding the $\mathrm{\bf {k}}(\k_{\rm x}=0, \k_{\rm y}=0,\k_{\rm z})$ modes respectively. We find that power is enhanced by $10-200 \%$ for $k \lesssim 0.1$ Mpc$^{-1}$. The reason is that for the light cone cube there is a gradual change in the mean brightness $\delta T_b$ with redshift. Large scale LOS modes with $k_x=k_y=0$ gain power because of this and hence affect the large $k$ modes in the spherically averaged power spectrum. We find that the coeval cubes are hardly affected because the mean $\delta T_b$ is similar for all slices. We also note that for the simulations studied in this paper, exclusion of the  $\mathrm{\bf {k}}(\k_{\rm x}=0, \k_{\rm y}=0,\k_{\rm z})$ modes is practically the same as the subtraction of the mean brightness temperature from each single frequency 21-cm map.

\end{document}